\documentclass{amsart}

\usepackage{amsthm,amssymb,amsmath}
\usepackage{bm}
\usepackage{bbm}
\usepackage{graphicx}
\usepackage{subfigure}
\usepackage{tikz}
\usepackage{array}
\usepackage{mathptmx}
\usepackage{multirow, makecell}
\usetikzlibrary{arrows,positioning,shapes.geometric}
\usetikzlibrary{matrix,chains,decorations.pathreplacing}

\usepackage{xcolor}
\usepackage{microtype}
\usepackage{booktabs}
\usepackage{url,diagbox}

\begin{document}

\title[Noninvasive fECG recovery]{Recovery of the fetal electrocardiogram for morphological analysis from two trans-abdominal channels via optimal shrinkage}

\author{Pei-Chun Su}
\address{Department of Mathematics, Duke University, Durham, NC, USA} 

\author{Stephen Miller}
\address{Department of Pediatrics, Division of Pediatric Cardiology, Duke University Medical Center, Durham, NC, USA} 

\author{Salim Idriss}
\address{Department of Pediatrics, Division of Pediatric Cardiology, Duke University Medical Center, Durham, NC, USA} 

\author{Piers Barker}
\address{Department of Pediatrics, Division of Pediatric Cardiology, Duke University Medical Center, Durham, NC, USA} 

\author{Hau-tieng Wu}

\address{Department of Mathematics and Department of Statistical Science, Duke University, Durham, NC, USA}

\begin{abstract}
We propose a novel algorithm to recover fetal electrocardiogram (ECG) for both the fetal heart rate analysis and morphological analysis of its waveform from two or three trans-abdominal maternal ECG channels. 
We design an algorithm based on the optimal-shrinkage under the wave-shape manifold model. 
For the fetal heart rate analysis, the algorithm is evaluated on publicly available database, 2013 PhyioNet/Computing in Cardiology Challenge, set A {(\texttt{CinC2013})}. 
For the morphological analysis, we {analyze \texttt{CinC2013} and another publicly available database, Non-Invasive Fetal ECG Arrhythmia Database (\texttt{nifeadb})}, and propose to simulate semi-real databases by mixing the MIT-BIH Normal Sinus Rhythm Database and MITDB Arrhythmia Database. 
For the fetal R peak detection, the proposed algorithm outperforms all algorithms under comparison. For the morphological analysis, the algorithm provides an encouraging result in recovery of the fetal ECG waveform, including PR, QT and ST intervals, even when the fetus has arrhythmia{, both in real and simulated databases}. 
To the best of our knowledge, this is the first work focusing on recovering the fetal ECG for morphological analysis from two or three channels with an algorithm potentially applicable for continuous fetal electrocardiographic monitoring, which creates the potential for long term monitoring purpose.

\end{abstract}

\maketitle

\section{Introduction} \label{sec:introduction}

Fetal cardiac arrhythmias are detected in at least 2\% of pregnancies \cite{Roland2011}. While many are transient, some sustained arrhythmias such as supraventricular tachycardia, ventricular tachycardia, and atrioventricular block may have significant consequences for the fetus and can result in fetal heart failure and demise. Therefore, detection and correct identification of fetal rhythm abnormalities is an important component of prenatal management {\cite{maeno2009fetal}}. Postnatal evaluation of the cardiac conduction system is performed by electrocardiography (ECG), with body-surface recording of cardiac electrical signal; however this is not possible in the fetus.
The first observation of fetal electrocardiograph (fECG) was reported by Cremer in 1906 \cite{Cremer:1906}, when he accidentally recorded a small portion of the fetal tracing while assessing the ECG of a pregnant woman. However, over the ensuing years, multiple methods to obtain a refined non-invasive recording of the fetal ECG have been applied {with some successes --} some positive outcomes have been reported {\cite{doi:10.1046/j.1471-0528.2003.02005.x,chia2005cardiac,CliffordSameni2011,Behar2016,lakhno2017use,behar2019noninvasive} and there are devices that have obtained FDA clearance, e.g., the Monica AN24 monitor from Monica Healthcare (Nottingham, UK) and the MERIDIAN M110 monitor from MindChild Medical (North Andover, MA, USA) \cite{clifford2014non}}. 
Fetal magnetocardiography has been developed, and successfully can provide morphology of the fetal ECG, including T wave morphology and QT interval \cite{STRASBURGER20081073,cuneo2013utero,Hornberger2008}. However, the technique is quite expensive and is very limited in availability, leading to infrequent use. 
Given these difficulties, fetal echocardiography has been used a surrogate to the actual fECG. Through a combination of 2D imaging, M-mode imaging and Doppler analysis, correct identification of premature beats, tachyarrhythmia, and bradycardia such as heart block can be made and can guide medical management \cite{Hornberger1331}.  
Despite the success of echocardiography, however, observation of the actual fECG signal would be very useful, especially for the diagnosis of conditions such as long QT syndrome, which may not result in obvious arrhythmias on echocardiography. Acquiring a morphologically refined fECG signal would assist in elucidating the origin and mechanisms of tachyarrhythmias and other conditions. Additionally, there has been some evidence that fetal ST analysis monitoring may help detect and alert fetal hypoxia, although this remains controversial \cite{Belfort_Saade:2015}.

In this paper, we focus on obtaining fECG from the ECG signal recorded from the mother's abdomen, where the sensor is close to the fetus so that the fECG signal can be recorded \cite{Jenkins1989,Sameni2010}. The signal is called {\em trans-abdominal maternal ECG} (ta-mECG). Importantly, the ta-mECG contains not only the fECG, but also the maternal cardiac activity, which we call the \textit{maternal abdominal ECG (mECG)}. 
Due to the existence of mECG, to obtain the fECG out of ta-mECG, we have to separate the fECG from the mECG. Since the fECG is often weaker than mECG, due to the inevitable noise, the fECG has a low signal-to-noise ratio (SNR). Moreover, due to the wide spectrum of ECG morphology and close heart rates of mother and fetus, the usual linear signal processing techniques do not work. All these facts make this biomedical signal processing challenging.

There have been many attempts to conquer this challenge in the past decades, and we can roughly classify those attempts into three categories -- (a) multiple ta-mECG channels, with or without one or more maternal thoracic ECG signals; (b) only single ta-mECG channel; (c) few (two or three) ta-mECG channels. 
Most attempts fall in category (a), and researchers apply algorithms like blind source separation (BSS) \cite{Lathauwer2000,Akhbari2013,DiMariaLiu2014,Varanini2014},
semi-BSS like periodic component analysis  \cite{SameniJutten2008,Haghpanahi2013,Akbari2015},
adaptive filtering like approaches \cite{Widrow1975,Graupe2008,Behar2014,Sameni2008,Niknazar2013,AndreottiRiedl2014,Panigrahy2017},
and others \cite{Ghaffari2015,Rodrigues2014,Lipponen2013}.
When there is only a single channel ta-mECG in category (b), researchers consider approaches like
template subtraction (TS) \cite{damen1982use,Cerutti1986,Oosterom1986, Vander1987, Martens2007,Ungureanu2007,Kanjilal1997,ChristovSimova2014}, 
time-frequency analysis \cite{Khamene2000,KarvounisTsipouras2007,CastilloMorales2013,Almeida2014,Lamesgin2015,SuWu2017}, 
sequential total variation \cite{LeeLee2016},
or state space reconstruction via lag map \cite{Richter1998,KarvounisTsipouras2009,Kotas2010}.
Only few papers focus on algorithms based on few (two or three) ta-mECG channels in category (c); for example, in \cite{LiFraschWu2017}, based on the dipole current model, the algorithm SAVER that combines the nonlinear time-frequency analysis and manifold learning achieve an accurate fHR estimation, and \cite{shnitzer2018recovering} applies the commutator of two diffusion operators to enhance the fetal ECG for the sake of fetal R peak detection. 
For {\em long term} monitoring, it is arguably better to have as few channels as possible. However, due to less information being available compared to having a multiple channel setup, the accuracy rate is expected to be lower when there is only one channel. It is shown in \cite{LiFraschWu2017,shnitzer2018recovering} that when there are two channels, we have more structure to use {compared with the single channel setup,} and can obtain significant improvement of the fHR prediction accuracy.

While there is rich literature for the fHR prediction, there are less literature focusing on the fECG recovery from the ta-mECG for the morphological analysis. There are some clinical studies comparing the fetal ST-segment change detected from the multi-channel ta-mECG recordings and the fetal ST analysis monitoring \cite{CliffordSameni2011}. {A scheme of evaluating} the accuracy of obtaining the fetal QT measurement from the ta-mECG is {proposed in} \cite{Behar2016}. 
{In \cite{lakhno2017use}, two cases of fetus with the second-degree fetal atrioventricular block is reported. In \cite{behar2019noninvasive}, the feasibility of diagnosis of fetal arrhythmias via multi-channel noninvasive fECG is carried out in the clinical setup with encouraging results.} 
In \cite{Niknazar2013}, the amplitude ratio of the extracted fECG by the extended state Kalman filtering is studied. In \cite{Andreotti2016}, researchers take a simulated data to evaluate the accuracy of estimating the fetal QT and the fetal T/QRS ratio from the ta-mECG. When there are multiple channels, the augmented time-sequenced adaptive filtering \cite{Vullings2018} is proposed to enhance the fECG for morphological analysis. When there are only one or two channels available, theoretically TS-based and adaptive filter-based algorithms \cite{Vullings2011} have the potential to recover the fECG for morphological analysis. In \cite{SuWu2017}, the nonlocal median technique is applied to enhance fECG morphology, but morphological analysis is not extensively discussed and quantified.

\subsection{Our contribution}
In this paper, we propose a novel algorithm to extract fECG for not only fHR estimation, but also for morphological analysis, when there are only {\em two or three} ta-mECG channels available. {A directly related algorithm is the singular value decomposition (SVD) approach  \cite{damen1982use,Oosterom1986, Vander1987, Kanjilal1997} in category (b), which was applied on a matrix of ta-mECG cardiac cycles centralized at maternal R peaks to estimate the mECG, so that the fetal cardiac cycles could be obtained by a direct subtraction of the top singular vector. However, the existence of fetal cardiac activity and noise biases the singular values and singular vectors, as discussed in Section \ref{Section:DipoleCurrentModel}. Moreover, disregarding singular vectors with smaller singular values also induces loss of morphology information. To solve these problems, the main novelty of this paper is introducing a robust mECG and fECG estimation, based on the currently developed {\em optimal shrinkage} (OS) theory for the matrix denoise.}
The OS theory comes from the fundamental random matrix theory when we handle the high dimensional dataset. With this metric, we obtain a better and more adaptive template for each cardiac cycle, and hence the morphology of the separated fECG is better reconstructed compared with the previous algorithms designed for few ta-mECG channels. 

The paper is organized as follows. In Section \ref{sec:Background}, we discuss the necessary mathematical models and backgrounds. In Section \ref{sec:algorithm}, the proposed algorithm is provided with implementation details. The material and evaluation are detailed in Sections \ref{Section:MaterialEvaluation} and \ref{Section:Evaluation}. The results are reported in Section \ref{Section:Result}. The paper is summarized in Section \ref{Section:Discussion} with a discussion.

\section{Model and Background}\label{sec:Background}

We summarize the necessary mathematical background in this section. Readers having interest in the algorithm can jump to Section \ref{sec:algorithm} without interruption.

\subsection{Dipole current model}\label{Section:DipoleCurrentModel}

Recall the dipole current model for the ECG signal \cite{Keener:1998}. In short, the model says that the recorded ECG signal is a projection of the {\em dipole current}, a 3-dim valued time series that is a surrogate of the overall cardiac electrophysiological (EP) activity, in different directions. Based on this model, the maternal ECG signal recorded from the maternal abdomen is 
\begin{equation}
E_{v_{m},m}(t)=v_{m}^{\top}D_{m}(t),
\end{equation}
where $D_{m}:\mathbb{R}\to \mathbb{R}^3$ represents the underlying dipole current of the maternal cardiac EP activity and $v_{m}\in S^2:=\{v\in \mathbb{R}^3|\,\|v\|_{\mathbb{R}^3}=1\}$ represents the projection direction, which reflects the relative location of the maternal heart and the lead placement.
Similarly, the fetal ECG signal recorded from the maternal abdomen is 
\begin{equation}
E_{v_{f},f}(t)=v_{f}^{\top}D_{f}(t),
\end{equation}
where $D_{f}:\mathbb{R}\to \mathbb{R}^3$ means the underlying dipole current of the fetal cardiac EP activity and $v_{f}\in S^2$ means the associated projection direction.
{We mention that some previous works in this area are based on the assumption that the maternal ECG has 3 statistically independent dimensions, while the fetal ECG is statistically 2-dimensional. However, as is discussed in \cite{sameni2006ica}, this assumption about the dimensionality of the fetal ECG might be true for the practical purpose, physiologically there is no reason for the dimensional difference between fetal and adult hearts.}
The ta-mECG signal is thus modeled as the linear summation of $E_{v_m,m}$ and $E_{v_f,f}$. 

In practice, the signal is sampled at the frequency $f_s$ Hz, and there might exist inevitable baseline wandering and noise. Suppose we place $J\in \mathbb{N}$ channels on the maternal abdomen. The $j$-th recorded ta-mECG, $j=1,\ldots,J$, is thus saved as $\mathbf{z}_j\in \mathbb{R}^N$ that satisfies 
\begin{equation}
\mathbf{z}_j(i)=v_{m,j}^{\top}D_{m}(i/f_s)+v_{f,j}^{\top}D_{f}(i/f_s)+T_j(i)+\xi_j(i), 
\end{equation}
where $i=1,\ldots,N$, $v_{m,j}, v_{f,j}\in S^{2}$ are the projection directions, $T_j$ is the baseline wandering, and $\xi_j$ is the noise.
In the ta-mECG setup, we have the following important observation regarding the lead setup. 
If two abdominal leads, $j\neq j'$, are far apart enough, which we can usually assume, then we know that $v_{f,j}$ and $v_{f,j'}$ are not collinear. On the other hand, since abdominal leads are far away from the maternal heart, {the mECG's, $v_{m,j}$ and $v_{m,j'}$, are  more collinear compared with the fECG's.} 

This observation allows us to consider the following linear combination scheme to enhance the fECG analysis.
Consider $\theta=(\theta_1,\ldots,\theta_J)\in S^{J-1}$, and define
\begin{equation}
\mathbf{z}_\theta(i)=\sum_{j=1}^J\theta_j \mathbf{z}_j(i) =v_{m,\theta}^{\top}D_{m}(i/f_s)+v_{f,\theta}^{\top}D_{f}(i/f_s)+\sum_{j=1}^J \theta_j T_j(i)+\sum_{j=1}^J \theta_j \xi_j(i), 
\end{equation}
where $v_{m,\theta}:=\sum_{j=1}^J \theta_j v_{m,j}$, and $v_{f,\theta}:=\sum_{j=1}^J \theta_j v_{f,j}$.
We view $\mathbf{z}_\theta$ as the ta-mECG with the projection direction for the mECG $v_{m,\theta}$ and the projection direction for the fECG $v_{f,\theta}$. By the above observation, since {$v_{m,j}$ and $v_{m,j'}$ are more collinear} when $j\neq j'$, by a proper design of the linear combination $\theta$, we are able to reach {a weakened mECG, $v_{m,\theta}$,}  in the linearly combined ta-mECG. On the other hand, since $v_{f,j}$ and $v_{f,j'}$ are not collinear, by a properly chosen $\theta$, the fECG could be enhanced in the new ta-mECG. In order to put this fact into practice, note that searching for the optimal $\theta$ such that {$v_{m,\theta}$ is small} is equivalent to searching for the optimal $\theta$ such that $v_{f,\theta}$ is strong relative to $v_{m,\theta}$.

\begin{figure}[ht]
\includegraphics[width=.99\textwidth]{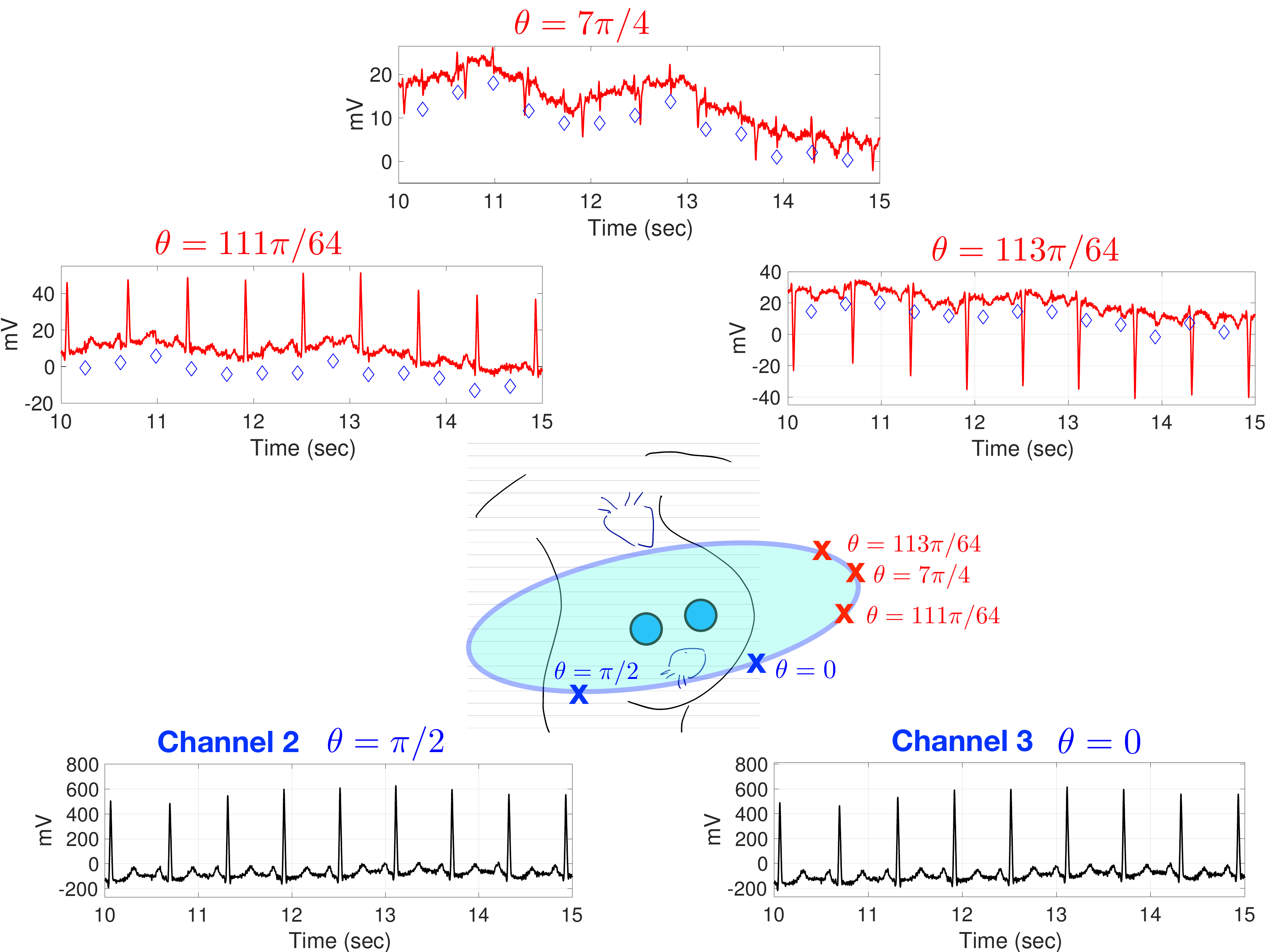}
\caption{An illustration of the dipole current model when $J=2$. The signal is from a59 of CinC2013 database, set A. The light blue circle indicates the $S^1$ representing linear combinations of Channel 2 and Channel 3. The Channel 2 and Channel 3 ta-mECG signals are shown in the bottom. Different linear combinations with different $\theta$'s are shown in the top. 
{Note that the insets with red signals have different y-axis minima and maxima. This decreases in the amplitude is caused by the cancelation so that the signal is gained to demonstrate fECG.}
The fetal R peaks provided by the experts are superimposed as blue diamonds. It is clear that when $\theta=(\cos(7\pi/4), \sin(7\pi/4))\in S^1$, the fECG is enhanced.}
\label{Figure:Dipole}
\end{figure}

Take $J=2$ as an example. If the two leads are far away enough, $v_{f,1}$ and $v_{f,2}$ are approximately orthonormal, and hence $v_{f,\theta}$ approximately is a projection. On the other hand, if {$v_{m,1}$ and $v_{m,2}$ are further assumed to be} $v_{m,1}\approx v_{m,2}$, which means that $\theta_1 v_{m,1}+\theta_2 v_{m,2}\approx  (\theta_1+\theta_2)v_{m,1} \approx (\theta_1+\theta_2) v_{m,2}$. Thus, if we choose $\theta_1=-\theta_2$ and if $v_{f,1}$ and $v_{f,2}$ are not collinear, the mECG is weakened and the {order} of fECG magnitude  is preserved in the linearly combined ta-mECG. See Figure \ref{Figure:Dipole} for an illustration when $J=2$ and the effect of canceling the mECG in the linear combination. Note that the y-axis range is different for insets demonstrating mECG canceling. It is clear that when $\theta_1=-\theta_2$ (that is, when we take $\theta=(1/\sqrt{2},-1/\sqrt{2})\in S^1$), the mECG is weakened so that the fECG is enhanced.

\subsection{Spike model and optimal shrinkage under the large $p$ large $n$ setup}\label{Section:TheorySpikeModel}

Matrix denoising problem is commonly encountered in several scientific fields. In general it asks if we are able to recover a $p\times n$ data matrix $X$ from its noisy version $\tilde{X}=X+N$, where $N$ is a $p\times n$ matrix modeling the noise. This seeming irrelevant problem plays a fundamental role in our proposed algorithm.

Obviously, without putting any condition, there is no way to solve this problem. Suppose the data matrix $X$ is of low rank, and $N$ is assumed to have independent and identical noise in each entry with zero mean, unit variance and finite fourth moment. Under these assumptions, an elegant solution based on the random matrix theory is proposed in \cite{svd_shrinkage2014} to recover the data matrix from $\tilde{X}$. Since $X$ is of low rank, a naive approach to recover $X$ is applying { SVD} combined with the noise level estimation. However, this naive approach is not optimal when $p$ and $n$ are both ``large''. Indeed, we cannot recover the singular values and singular vectors without any bias under this setup. Mathematically it is quantified in the following way. When $p=p(n)$ and $p(n)/n\to \beta>0$ when $n\to \infty$, asymptotically the singular vectors of $\tilde{X}$ are deviated from the singular vectors of $X$ and we do not have an unbiased estimator. Usually, the above setup is understood as the {\em spike model}. To denoise the matrix under the spike model, we need to take this bias issue into account. In \cite{svd_shrinkage2014}, the authors propose to correct the singular values to achieve a better denoised matrix, where ``better'' is determined by the loss function determined by the user. For example, we can define the loss function to be $L_{op}(\hat{X}|X):=\|X-\hat{X}\|_{op}$, and find the optimal $\hat{X}$ from $\tilde{X}$ to minimize $L(\hat{X}|X)$. Denote the singular value decomposition of $\tilde{X}$ as $U\Lambda V^{\top}$, where $U\in O(p)$ and $V\in O(n)$ are the left and right singular vector matrices, $\Lambda\in \mathbb{R}^{p\times n}$ contains singular values $\sigma_1\geq\ldots\geq \sigma_{\min\{p,n\}}$ in the diagonal entries.
If the operator norm is considered as the loss function, $L_{op}$ and $p/n\leq 1$, then it is shown in \cite{svd_shrinkage2014} that the optimal denoised matrix is given by
\begin{equation}\label{FormulaSVDDenoise}
\hat{X}=U\Lambda^*V^{\top},
\end{equation}
where $\Lambda^*\in \mathbb{R}^{p\times n}$ so that $\Lambda^*(i,i)=\eta^*(\sigma_i)$ for $i=1,\ldots,p$ and $\eta^*$ is the asymptotically optimal shrinker defined on $\mathbb{R}_+$ given by
\begin{equation}\label{Definition OS eta}
\eta^*(\lambda) = \frac{1}{\sqrt{2}} \sqrt{\lambda^2-\beta-1+\sqrt{(\lambda^2-\beta-1)^2-4\beta}}\,
\end{equation}
when $\lambda\geq 1+\sqrt{\beta}$ and $\eta^*(\lambda)=0$ otherwise. 

The above setup fits our needs in the fECG extraction algorithm. Take the mECG morphology as an example. It is well known that the R peak amplitude varies due to the impedance changes caused by respiration, and the QT interval changes nonlinearly with related to the RR interval. Therefore, the mECG morphology, including P-QRS-T waveforms, varies from time to time. However, it does follow some physiological rule. This physiological fact says that if we align R peaks of mECG segments associated with maternal cardiac cycles, then we only need few parameters to well parametrize these mECG segments. Mathematically, this fact says that the matrix $X=[x_1,\ldots,x_n]\in \mathbb{R}^{p\times n}$ containing $n$ mECG segments, $x_1,\ldots,x_n$, each of which is of length $p$, is of low rank. For the ta-mECG we have interest, it is contaminated by noise and the existence of fECG. Therefore, the matrix $\tilde{X}=[\tilde{x}_1,\ldots,\tilde{x}_n]\in \mathbb{R}^{p\times n}$ containing all segments containing maternal cardiac cycles can be modeled as a noisy matrix that
\begin{equation}
\tilde{X}=X+N,
\end{equation}
where $X$ contains purely mECG segments, and $N$ contains noise and fECG. 
{Therefore, we could recover $X$ from $\tilde{X}$ by \eqref{FormulaSVDDenoise}. From the signal processing perspective, unlike the commonly applied Fourier-based filtering technique, this OS approach is an ``adaptive'' bandpass filtering scheme, where the adaptivity comes from finding a basis adaptive to the dataset via SVD. In practice, this OS approach can be applied to analyze as few as a single ta-mECG channel, or several ta-mECG channels. But in the current work, we focus on the single ta-mECG setup, which comes from a linear combination of few ta-mECG channels.}
Here, we implicitly make an assumption that the mHR and the fHR do not couple; that is, we assume that the fetal heart cycles can happen at any phase of one maternal heart cycle, and hence the randomness. When the mHR and fHR are coupled, we need more information and it is out of the scope of this paper.

\section{Proposed Algorithm}\label{sec:algorithm}

\begin{figure}[ht]
\includegraphics[width=.98\textwidth]{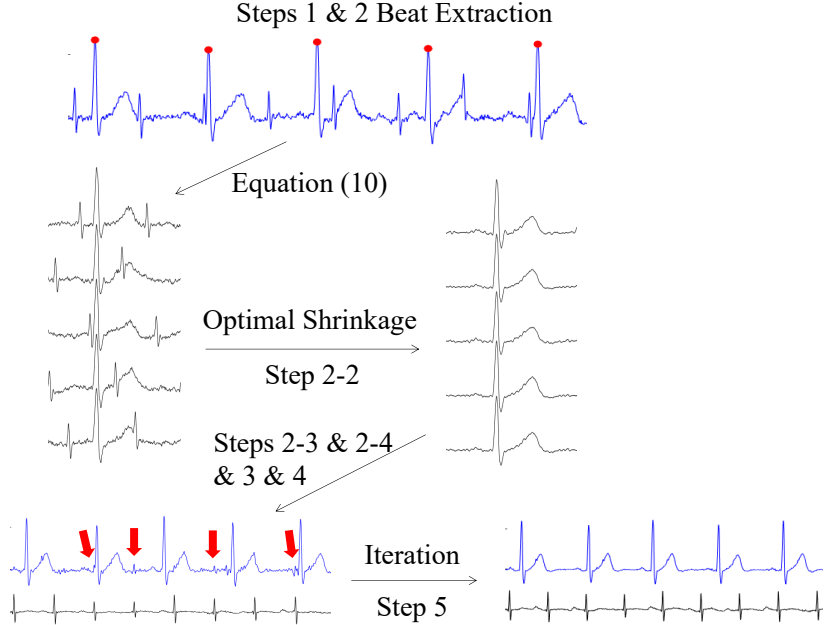}
\caption{
The illustration of the proposed algorithm. Top: the optimal linearly combined ta-mECG with the maternal R peaks superimposed as read circles. The maternal R peaks estimated by the de-shape short time Fourier transform algorithm {and the beat tracking algorithm} proposed in \cite{SuWu2017}.
Middle left: for each maternal R peak, the associated ta-mECG cycles {are stacked together with the associated maternal R peaks aligned. Only five beats are shown to enhance the visualization.} 
Middle Right: by applying the optimal shrinkage on the stacked segments, the noise and fECG are suppressed.
{Bottom left:} The mECG signal is reconstructed and the rough fECG is obtained by subtracting the reconstructed mECG (black line) from the ta-mECG. {Bottom right:
We can further repeat the R peak detection, optimal shrinkage, and median on the rough fECG to enhance the fECG morphology. Clearly, the small remaining fECG in the estimated mECG indicated by the orange arrows on the bottom left subplot is alleviated in the estimated mECG shown on the bottom right subplot. As a result, the fECG morphology is better recovered.
}
}
\label{Figure:WholeAlgorithm}
\end{figure}

The proposed algorithm falls in the category of {\em single channel blind source separation} (scBSS) algorithm for the general purpose. Specifically, the algorithm separates the fetal ECG and maternal ECG out of the recorded ta-mECG. 
There are three main steps (Steps 2-1, 2-2 and 2-3 below) in addition to the standard pre-processing step (Step 1 below). First, estimate the maternal heart rate based on the nonlinear-type time-frequency analysis called de-shape short-time Fourier transform (dsSTFT) \cite{SuWu2017,LiFraschWu2017}. Then, divide the ta-mECG into pieces so that each piece contains one maternal cardiac cycle. Second, design a {\em metric} to compare pieces. The novelty of the proposed algorithm is utilizing the {\em optimal shrinkage} (OS) tool that is immune to information not related to the maternal cardiac cycles, like the fetal cardiac activities and noise. With this metric, for each piece, we find other pieces with similar maternal cardiac cycles. 
Finally, the median of all similar maternal cardiac cycles is evaluated, which recovers the mECG. By repeating these three steps for all pieces, the fECG is recovered. 
See Figure \ref{Figure:WholeAlgorithm} for a summary illustration of the proposed algorithm. Below we detail the algorithm step by step for the reproducibility purpose.

\subsection{Input data}

Fix a subject. If the sampling rate of the data from one subject is less than $1000$ Hz, to enhance the R peak alignment needed in the following steps, the signal is upsampled to $1000$ Hz \cite{Laguna2000}. Thus we assume below that all signals are sampled at $f_s=1000$ Hz.
Suppose all recordings have $J\in \mathbb{N}$ simultaneously recorded ta-mECG channels. Denote the $j$-th channel $\mathbf{x}^0_{j} \in \mathbb{R}^{N}$, where $j=1,\ldots,J$ and $N\in\mathbb{N}$ is the number of samples of each channel; that is, the recording is over an interval of $N/f_s$ seconds.

\subsection{Step 1: Preprocessing and linear combination}

The Butterworth low-pass filter of order 5 with the cut-off frequency of 100 Hz and a notch filter with the notch centered at 60Hz are applied on each ta-mECG signal to remove high frequency noises and the powerline interference noise. 
To remove the baseline wandering, which corresponds to low frequency noise caused by a variety of sources including respiration, body movements, and poor electrode contact, we apply the following two stage moving window median filter with window size $200$ and $600$ ms. This two stage scheme presents higher signal to noise ratio (SNR) in \cite{Sameni2008} and shows the ability to preserve fECG morphology in \cite{Panigrahy2017}. 
Let $\mathbf{D}_S$ and $\mathbf{D}_L$ denote the operator of median filter with short and long window size respectively. The baseline wandering of $\mathbf{x}^0_{j}$, $j=1,\ldots,J$, is estimated by $\mathbf{D}_L\mathbf{D}_S\mathbf{x}^0_{j}$, and hence the detrended signal $\mathbf{x}_{j}= \mathbf{x}^0_j-\mathbf{D}_L\mathbf{D}_S\mathbf{x}^0_{j}$.

Following the key idea described in \cite{LiFraschWu2017}, produce a variety of ta-mECG signals constructed by different linear combinations of $\mathbf{x}_{j}$, $j=1,\ldots,J$. Consider a finite subset $\Theta\subset S^{J-1}$, and for each $\boldsymbol\theta=(\theta_1,\ldots,\theta_d)\in \Theta$, construct a linear combination: 
\begin{equation}
\mathbf{z}_{\boldsymbol\theta} = \sum_{j=1}^J{\theta_j}\mathbf{x}_{j} \,.
\end{equation}
This construction is based on the dipole current model of ECG signal \cite{Keener:1998} discussed in Section \ref{Section:DipoleCurrentModel}. 
In the following, when $J=2$, we set $\Theta = \{(\theta_1,\theta_2)\}$, where $\theta_1\in \{-\frac{6}{7}, -\frac{5}{7},\ldots,1\}$. 
Note that when there is only one channel, $S^0=\{-1,1\}\subset \mathbb{R}$, and there is no linear combination.

\subsection{Step 2-1: Estimate maternal R peaks}

In this step we estimate the maternal R peaks.  
Due to the existence of fetal ECG signal, the traditional R peak detection algorithms might fail. We apply the idea proposed in \cite{SuWu2017} to obtain the R peaks by taking the beat tracking algorithm into account. We refer readers with interest to \cite{SuWu2017,LiFraschWu2017} for the algorithm details, and \cite{LinSuWu2018} for the underlying theory.
Denote the collected timestamps of estimated maternal R peak locations of  {the linearly combined ta-mECG $\mathbf{z}_\theta$} as 
\begin{equation}
\mathbf{R}_{\theta}^{(m)} = \{R_{\theta}^{(m)}(i)\}_{i=1}^{n_{\theta}^{(m)}}\,,
\end{equation}
where $R_{\theta}^{(m)}(i)$ is the timestamp of the $i$-th detected maternal R peak and $n^{(m)}_{\theta} \in \mathbb{N}$ is the total number of estimated maternal R peaks. 
As is mentioned in our model, in some linear combinations, it is possible that {$v_{m,\theta}$ is too small so that} the amplitudes of $v_{m,\theta}$ and $v_{f,\theta}$ are comparable, {or even $v_{f,\theta}$ is dominant,} which will result in wrong estimation of ${\bf{R}}_{\theta}^{(m)}$. To solve this problem, we apply the fusion method \cite{Yu2016FusionOD}. { When $J=2$, out of 14 linear combinations, we collect 5 linear combinations so that the associated detected R peaks have the smallest heart rate variability. Then we apply a voting procedure to determine the final maternal R peaks for all linear combinations.} For simplicity, we still denote the final maternal R peaks for each linear combination as ${\bf{R}}_{\theta}^{(m)}$.

\subsection{Step 2-2. Estimate maternal cardiac cycles by optimal shrinkage}
 
{\bf $\bullet$ Construct mECG templates.}
Fix a linear combination $\theta$, and the $i$-th maternal cardiac cycle determined by the $i$-th R peak. 
Denote $w_{\theta}\in \mathbb{N}$ to be the rounding number of the {$95\%$ quantile} of R to R intervals of $\mathbf{z}_{\theta}$. 
Denote the corresponding ECG segment over the cardiac cycle as
\begin{equation}\label{Equation:ConstructSegments}
\mathbf{s}^{(m)}_{\theta,i} :=  \left[\mathbf{z}_{\theta}\left(R_{\theta}^{(m)}(i)-\Big\lceil \frac{3w_{\theta}}{8}\Big\rceil\right),\ldots, \mathbf{z}_{\theta}\left(R_{\theta}^{(m)}(i)+\Big\lceil\frac{5w_{\theta}}{8}\Big\rceil\right)\right]^\top\in \mathbb{R}^{p_{\theta}}\,,
\end{equation}
where $\lceil x\rceil$ is the smallest integer larger than $x>0$ and $p_{\theta}:=\Big\lceil\frac{3w_{\theta}}{8}\Big\rceil+\Big\lceil\frac{5w_{\theta}}{8}\Big\rceil+1$. 
Here, the values $3/8$ and $5/8$ are set based on the knowledge of PR and QT duration \cite{bioengineering3040026}, so that the whole P-QRS-T waveform is covered.
Build up a library for the mECG template, denoted as
$\mathcal{L}^{(m)}_{\theta}:=\{\mathbf{s}^{(m)}_{\theta,i}\}_{i=1}^{n_{\theta}^{(m)}}$.
\newline\newline
{\bf $\bullet$ Remove nuisance variables by optimal shrinkage.}
Construct a data matrix $\mathbf{S}_{\theta}$ of size $p_{\theta}\times {n_{\theta}^{(m)}}$ consisting of all segments of $\mathcal{L}^{(m)}_{\theta}$ in the columns. To obtain maternal cardiac cycles from $\mathcal{L}^{(m)}_{\theta}$, we need to reduce the influence of the fECG and noises. Based on the assumption mentioned in the manifold model, and the low-rank assumption of the mECG inside the data matrix $\mathbf{S}_{\theta}$, we
apply the optimal shrinkage \cite{svd_shrinkage2014} as we discussed in 2.2.. Suppose $\frac{p_{\theta}}{n^{(m)}_{\theta}}\leq 1$.
Denote the SVD of the data matrix $\mathbf{S}_{\theta}$ as 
\begin{equation}
\mathbf{S}_{\theta} = \sum^{r}_{i=1}\lambda_i\mathbf{u}_i \mathbf{v}'_i\,,
\end{equation}
where $r$ is the matrix rank, $\mathbf{u}_i$ and $\mathbf{v}'_i$ are the $i$-th left and right singular vector corresponding to the singular
value $\lambda_i$, and $\lambda_1\geq \lambda_2\geq\ldots\geq \lambda_{p_{\theta}}$. {Since equation \eqref{Definition OS eta} is constructed under the assumption of unit variance of noise level, we normalize $\mathbf{S}_{\theta}$ with noise level estimated by

\begin{equation}
\varsigma_{\theta} := C_{\theta} \cdot \sqrt{\frac{1}{n^{(m)}_{\theta} \cdot p_{\theta}}\sum^{n^{(m)}_{\theta}}_{i=1}\sum^{p_{\theta}}_{k=1}(\mathbf{s}^{(m)}_{\theta,i}(k) - \bar{\mathbf{s}}^{(m)}_{\theta}(k))^2}\,,
\end{equation}
where $C_{\theta}$ is a constant chosen by the user, 
\begin{equation}
\bar{\mathbf{s}}_{\theta}^{(m)}(k) =\text{median}\Big\{\mathbf{s}^{(m)}_{\theta,1}(k),\ldots,\mathbf{s}^{(m)}_{\theta,n^{(m)}_{\theta}}(k)\Big\}
\end{equation}
and $k=1,\ldots,p_{\theta}$.
We set $C_{\theta} = 1.5$ empirically to avoid underestimating the noise level. 
Then, we apply the OS as equation \eqref{Definition OS eta} as
\begin{equation}\label{CriticalStepOSStep}
\tilde{\mathbf{S}}^{\eta^*}_{\theta} =\varsigma_{\theta} \sum^{r}_{i=1}\eta^*\Big(\frac{\lambda_i}{\varsigma_{\theta}}\Big)\mathbf{u}_i\mathbf{v}'_i\,.
\end{equation}
The columns of $\tilde{\mathbf{S}}^{\eta^*}_{\theta}$ are the estimated mECG of each $\mathbf{s}^{(m)}_{\theta,i}$, denoted as $\tilde{\mathbf{s}}^{(m)}_{\theta,i}$. 
}

\subsection{Step 2-3. Maternal ECG recovery}

After obtaining all estimated maternal cardiac cycles, we reconstruct the estimated mECG signal from $\big\{\tilde{\mathbf{s}}^{(m)}_{\theta,i}\big\}_{i=1}^{n_{\theta}^{(m)}}$ by the standard stitching approach \cite{SuWu2017}.  Denote the estimated mECG from $\mathbf{z}_{\theta}$ as $\tilde{\mathbf{z}}_{\theta}^m$.

\subsection{Step 3. Channel Selection by Signal Quality Index}
For each linear combination $\mathbf{z}_{\theta}$ with the estimated mECG $\tilde{\mathbf{z}}^m_{\theta}$, we obtain the {\em rough fECG} by a simple subtraction:
\begin{equation}\label{RemovalmECG}
\mathbf{z}^{rf}_{\theta} = \mathbf{z}_{\theta} - \tilde{\mathbf{z}}^m_{\theta} 
\end{equation}
Denote $\{\mathbf{z}^{rf}_{\theta}\}_{\theta\in \Theta}$ to be the collection of rough fECG signals estimated from each linear combination.  We apply the bSQI \cite{JohnsonBehar2015} to determine the signal quality index (SQI) for each rough fECG. The optimal linear combination out of $\Theta$, denoted by $\theta^*$, is selected as the one that has the highest SQI. From now on, we fix to $\theta^*$, and hence $\tilde{\mathbf{z}}_{\theta^*}^{rf}$ to be the rough fECG of the given recording.

\subsection{Step 4. Get the fIHR}

We repeat Step 2-1 on $\tilde{\mathbf{z}}_{\theta^*}^{rf}$ to acquire the fetal heart beat locations. 

{
\subsection{Step 5. Enhance the mECG and rfECG by iteration}

Note that to carry out the OS in Step 2-2, we assume that the summation of fECG and noise is a realization of an independent and identical random variable of zero mean, unit variance and finite fourth moment. Recall the underlying theory in Section \ref{Section:TheorySpikeModel}. However, this assumption might not be fully satisfies for the fECG. 
We thus enhance the whole process by iterating Steps 2-2, 2-3 and 4. 

Denote the total number of detected fetal R peaks in $\tilde{\mathbf{z}}_{\theta^*}^{rf}$ by ${n_{\theta^*}^{(f)}}$. Then repeat Step 2-2 to obtain a set of denoised fetal cardiac cycles, denoted as $\tilde{\mathcal{L}}^{(f)}_{\theta^*}:=\{\tilde{\mathbf{s}}^{(f)}_{\theta^*,i}\}_{i=1}^{n_{\theta^*}^{(f)}}$, where $\tilde{\mathbf{s}}^{(f)}_{\theta^*,i}$ is the $i$-th denoised fetal cardiac cycle.
Reconstruct the fECG signal $\tilde{\mathbf{z}}_{\theta^*}^{f}$ from $\tilde{\mathcal{L}}^{(f)}_{\theta^*}$ by the standard stitching approach \cite{SuWu2017}.

Next, improve the mECG estimation by repeating Steps 2-2 and 2-3 on $\tilde{\mathbf{z}}_{\theta^*}^{rm} :=  {\mathbf{z}}_{\theta^*} - \tilde{\mathbf{z}}^f_{\theta^*}$. Note that the fECG component is now suppressed in $\tilde{\mathbf{z}}_{\theta^*}^{rm}$, so the noise assumption for the OS is better satisfied and we can better recover the mECG, denoted as $\hat{\mathbf{z}}_{\theta^*}^{m}$, and hence the rough fECG, denoted as $\hat{\mathbf{z}}_{\theta^*}^{rf}$.  
Finally, we repeat Step 4 on $\hat{\mathbf{z}}_{\theta^*}^{rf}$ to acquire the fetal heart beat locations and repeat Step 2-2 on $\hat{\mathbf{z}}_{\theta^*}^{rf}$ to obtain our proposed estimator of the fECG, denoted as $\hat{\mathbf{z}}_{\theta^*}^{f}$. 
}

\subsection{(Optional) Enhance the fECG signal {by the nonlocal median}}

{ 
When the fetus is normal sinus rhythmic, we may further improve the fECG morphology by applying the nonlocal median. Denote $\hat{\mathcal{L}}^{(f)}_{\theta^*}:=\{\hat{\mathbf{s}}^{(f)}_{\theta^*,i}\}_{i=1}^{n_{\theta^*}^{(f)}}$ to be the set of fetal cardiac cycles in $\hat{\mathbf{z}}_{\theta^*}^{f}$.
Take $k_m\in\mathbb{N}$, and denote $\mathcal{N}_{\theta^*,i} = \{\hat{\mathbf{s}}^{(f)}_{\theta^*,i_\ell}\}_{\ell=1}^{k_m}$ to be those $\hat{\mathbf{s}}^{(f)}_{\theta^*,i_\ell}\in \hat{\mathcal{L}}^{(f)}_{\theta^*}$ that have the most similar R-R intervals compared with that of $\hat{\mathbf{s}}^{(f)}_{\theta^*,i}$.
With $\mathcal{N}_{\theta^*,i}$, we take the entry-wise median of these segments as the estimation of corresponding fetal P-QRS-T waveform; that is,
the $i$-th fetal P-QRS-T waveform in $\mathbf{z}_{\theta^*}$ is estimated by
\begin{align}
\hat{\mathbf{s}}_{\theta^*,i}^{(cf)}:=\text{median}\{\hat{\mathbf{s}}^{(f)}_{\theta^*,i}, \hat{\mathbf{s}}^{(f)}_{\theta^*,i_1},\ldots,\hat{\mathbf{s}}^{(f)}_{\theta^*,i_{k_m}}\}\,.
\end{align}
Since the median filter is not applied to consecutive cycles, it is called the nonlocal median filter.
The nonlocal median filtered fECG is determined by stitching $\{\hat{\mathbf{s}}_{\theta^*,i}^{(cf)}\}_{i=1}^{n_{\theta^*}^{(f)}}$. 
}

\section{Material}\label{Section:MaterialEvaluation}

To validate the proposed algorithm in not only detecting the fetal heart rate, but also recovering the fetal ECG morphology, we use the following databases.

\subsection{Real database}\label{materials}

The first real database is the database {\em 2013 PhysioNet/Computing in Cardiology Challenge} \cite{Goldberger_Amaral_Glass_Hausdorff_Ivanov_Mark_Mietus_Moody_Peng_Stanley:2000}\footnote{\url{https://physionet.org/challenge/2013/#data-sets}}, abbreviated as \texttt{CinC2013}. We focus on the set A, which is composed of 75 recordings with the provided fetal R peak annotations. Each recording includes four ta-mECG channels that were obtained from multiple sources resampled at the sampling rate 1000 Hz and last for 1 minute duration. There is no publicly available information about where the leads are placed on the maternal abdomen. Case a54 is discarded based on the suggestion in \cite{AndreottiRiedl2014} since it was discarded by the Challenge organizers. We focus on the remaining 74 recordings.

{To demonstrate the performance of recovering fECG morphology, particularly when the fetus is arrhythmic, we also consider the {\em Non-Invasive Fetal ECG Arrhythmia Database} (\texttt{nifeadb}) \cite{Goldberger_Amaral_Glass_Hausdorff_Ivanov_Mark_Mietus_Moody_Peng_Stanley:2000}\footnote{\url{https://physionet.org/physiobank/database/nifeadb/}}. 
The \texttt{nifeadb} contains 12 fetal arrhythmias recordings and 14 normal rhythm recordings. Each recording has one maternal thoracic signals and four or five ta-mECG signals, with the 500 Hz or 1000 Hz sampling rate. 
The diagnosis information and more details can be found in \cite{behar2019noninvasive}.}

\subsection{Simulated database}

Since there is no gold standard recording of fetal ECG morphology available in any publicly available database, including \texttt{CinC2013}, to validate that our proposed algorithm has the capacity to recover the morphology of fECG from the ta-mECG, we evaluate our algorithm on a set of {semi-real} simulated ta-mECG data. We consider {two} databases. 

{The first one is the MITDB arrhythmia database \url{https://www.physionet.org/physiobank/database/mitdb/}, abbreviated as \texttt{MITDB}. 
This dataset contains 48 half-hour excerpts of two-channel ambulatory ECG recordings, obtained from 47 subjects studied in the Boston's Beth Israel Hospital Arrhythmia Laboratory between 1975 and 1979. 
Twenty-three recordings were chosen at random from a set of 4,000 24-hour ambulatory ECG recordings collected from 
a mixed population of inpatients (about 60\%) and outpatients (about 40\%) at Boston's Beth Israel Hospital; 
the remaining 25 recordings were selected from the same set to include less common but clinically significant arrhythmias 
that is not well-represented in a small random sample. 
Each subject has 2 channels with the sampling frequency 360 Hz and the 11-bit resolution over a 10 mV range. The R peak annotations are provided.}
The {second} one is the Physikalisch-Technische Bundesanstalt (PTB) Database \url{https://physionet.org/physiobank/database/ptbdb/}, 
abbreviated as \texttt{PTBDB}. The database contains 549 records from 290 subjects aged 17 to 87 with the mean age 57.2. Each subject is represented by one to five records. Each record includes 15 simultaneously measured signals: the conventional 12 leads (I, II, III, AVR, AVL, AVF, V1, V2, V3, V4, V5, V6) together with the Frank lead ECGs (Vx, Vy, Vz). 
Each signal is digitized with the sampling frequency 1000 Hz and with the 16 bit resolution over a range of $\pm 16.384$ mV. 
Out of 290 subjects, 216 subjects have cardiological disorders, 52 subjects are healthy, and 22 subjects do not have available clinical summary.

To create a mECG in the simulated ta-mECG, we take Vx, Vy, and Vz recordings from healthy subjects in \texttt{PTBDB}, denoted as $V_x(t),V_y(t),V_z(t)$ at time $t\in \mathbb{R}$, which are 115 seconds in total. We view the triplet $(V_x(t),V_y(t),V_z(t))$ as the maternal vectocardiogram (VCG).
We represent the project direction of maternal VCG by a pair of angles, $\theta_{xy}$ and $\theta_{z}$. The simulated mECG is then created by
\begin{equation}
\text{mECG}(t)=(V_x(t) \cdot \cos{\theta_{xy}} + V_y(t) \cdot \sin{\theta_{xy}})\cdot \cos{\theta_{z}} + V_z(t) \cdot \sin{\theta_{z}} \,.
\end{equation} 
Each subject has recordings created with two pairs of directions given by $(\theta_{xy},\theta_z) = (\frac{\pi}{4},\frac{\pi}{4})$ and $(\theta_{xy},\theta_z) = {(\frac{\pi}{5},\frac{3\pi}{10})}$. In total, {40} mECGs of healthy subjects are generated. Note that in this simulation, we assume that the mECG is arrhythmia-free.
{
The {\em simulated fECG's} of healthy fetus are created from another 40 healthy subjects in the \texttt{PTBDB}, where $114$ seconds of the V2 and V4 recordings are taken. 
The {\em simulated fECG's} of fetus with arrhythmia are created by taking the first $114$ seconds from the first and second channel of subjects in the \texttt{MITDB}.} The simulated mECG and simulated fECG come from different subjects.
Both healthy and arrhythmic signals are resampled at $500$ Hz.
As a result, the simulated fECG has the same number of data points as the simulated mECG, and has about double the heart rate compared with the simulated mECG if we consider both parts sampled by $1000$ Hz. The amplitude of the simulated fECG is normalized to the same level of simulated mECG and then multiplied by $0<r<1$ to make the amplitude smaller then the mECG, consistent with the usual situation of real ta-mECG signals.
We generate {40} simulated healthy and arrhythmic fECGs.
The clean simulated ta-mECG is generated by directly summing simulated mECG and fECG. The final simulated ta-mECG is generated by adding a Gaussian white noise to the clean simulated ta-mECG according to the assigned signal-to-noise (SNR) ratio.
As a result, we acquire {$40$} recordings of {$57$} seconds simulated ta-mECG signals with the sampling rate $1000$ Hz.

{
We mention that it is likely true that fECG morphology is less diverse compared with adult ECG morphology, in that the QRS duration is narrower, and QRS abnormalities such as bundle branch block, aberrant conduction, polymorphic ventricular ectopy or ventricular tachycardia are less common in the fetus. 
However, it does not mean that fetus does not have rate-related aberrancy or aberrant conduction following premature atrial beats, and it is not clear how often ventricular ectopy/tachycardia occurs and if this is monomorphic or polymorphic. See, for example, \cite{Zhao2008,Hornberger2008}, for a report of fetal ventricular ectopy in the setting of complete heart block with and without structural heart disease.
If the proposed algorithm could perform well on this more challenging semi-real ta-mECG signal for the morphologic analysis, then it would be reasonable to surmise that it would excel at fetal ECG morphology analysis.
}

\section{Evaluations}\label{Section:Evaluation}
\subsection{Evaluation of R peak Estimation.}

To evaluate the fetal R peak detection performance, a detected R peak is compared with the provided annotations by beat-to-beat comparisons with a matching window of $50$ ms \cite{AndreottiRiedl2014}; that is, a detected R peak is a true positive if it is close to a true R peak within a $50$ ms deviation. We report the $\textup{F}_1$ score 
\begin{equation}\label{Definition F1}
\textup{F}_1: = \frac{2\texttt{TP}}{2\texttt{TP}+\texttt{FN}+\texttt{FP}}\,,
\end{equation}
where \texttt{TP} is the true positive rate (correctly detected R peak), \texttt{FP} is the false positive rate (falsely detected R peak), and \texttt{FN} is the false negative rate (the existing R peak that is not detected). 
We also compute the {\em mean absolute error} (MAE), which is defined by
\begin{equation}
\textup{MAE} = \frac{1}{n_{\texttt{TP}}} \sum_{i=1}^{n_{\texttt{TP}}} |t_i^f - \tilde{t}_i^f|\,,
\end{equation}
where $n_{\texttt{TP}}$ is the number of \texttt{TP} detected R peaks, and $\tilde{t}_i^f$ and $t_i^f$ are the timestamps of the $i$-th \texttt{TP} detected R peak and the associated true R peak. Here, we follow the common approach \cite{Andreotti2016} to consider only \texttt{TP} R peaks to avoid the evaluation dependence on the R peak detection accuracy.
For \texttt{CinC2013}, we report the mean and medium of $\textup{F}_1$ and MAE for each channel combination over all recordings. For each recording, we evaluate the $\alpha$-quantile, where $\alpha\in[0,1]$, of all $\textup{F}_1$'s (respectively MAE's) of all channel combinations, denoted as $\textup{F}_1(\alpha)$ (respectively MAE($\alpha$)). Then we report the mean and median of the $\textup{F}_1(\alpha)$ (respectively MAE($\alpha$)) over all recordings. The practical meaning of $\textup{F}_1(1)$ (respectively MAE(1)) is the best possible result we can obtain from any combination of channels for a single recording. 

\subsection{Comparison with other algorithms}

When we have more than one channel ta-mECG signals, the fECG decomposition problem falls in the category of the blind source separation (BSS) and its variations \cite{SameniJutten2008,Haghpanahi2013,Akbari2015}. 
For the comparison purpose, we show the independent component analysis (ICA). There are several approaches to implement ICA, for example, the joint approximation diagonalization of eigen-matrices (JADE), and symmetric and deflationary FAST-ICA approaches. In \cite{Andreotti2016}, it is shown that JADE produced slightly better results. Thus, in this work, we only show the JADE result for the ICA approach. 
We apply the benchmark codes of JADE implementation of ICA provided in \url{http://www.fecgsyn.com}, and denote these method as $\text{BSS}_{\text{ICA}}$\footnote{Here we follow the nomination proposed in \cite{Andreotti2016}.}. {Another commonly applied method is principle component analysis (PCA). However, as is indicated in \cite[Section 3.3.2]{Behar2014thesis} that there is no reason for the mECG and fECG to be orthogonal in the observation space, PCA as a BSS approach might not be a suitable approach, so we do not consider it.}

A critical step in the BSS approach is identifying the decomposed signal that contains the maternal or fetal ECG \cite{AndreottiRiedl2014}. 
Since there are only two (or three) decomposed signals when we apply $\text{BSS}_{\text{ICA}}$ to two (or three) ta-mECG channels, it is not feasible to select the optimal channel. For the $\text{BSS}_{\text{ICA}}$ algorithm, we thus take the ground truth annotation to select the optimal channel that is more likely to be the fECG, and report the detected R peaks from this detected channel. 
%
We emphasize that we {\em do not} take the ground truth annotation into account in any other algorithms, particularly our proposed algorithm.

When the maternal thoracic-lead ECG signal (mtECG) is available, we can apply the adaptive filter (AF) idea to remove the mECG from the (possibly single channel) ta-mECG, where the maternal thoracic ECG signal (mtECG) is viewed as the {\em reference} channel. For example, the least mean square (LMS) \cite{Widrow1975} or the recursive least square (RLS) \cite{Behar2014} and its variations, like the echo state neural network (ESN) \cite{Behar2014}, blind adaptive filtering \cite{Graupe2008},
extended Kalman filter (EKF) \cite{Sameni2008}, etc.
If the mtECG and the mECG in the ta-mECG are linearly related, the LMS or RLS helps to extract the fECG by removing the maternal cardiac activity in the ta-mECG. If the relationship is nonlinear, ESN could help.  
In our setup, we cannot get the mtECG, so these AF-based algorithms cannot be directly applied. However, recall that we are able to accurately estimate the mECG in the ta-mECG \cite{SuWu2017,LiFraschWu2017}. Therefore, we could view the estimated mECG signal as the reference channel. We mention that this idea is also considered in \cite{Rodrigues2014}. 
Based on this idea, we consider the {\em modified} AF-based algorithms proposed in \cite{LiFraschWu2017}. We replace the direct subtraction step in (\ref{RemovalmECG}) by the LMS, ESN or EKF, by taking the estimated mECG as the reference channel to get the rough fECG. Note that since other steps are not changed, when there are more than two channels, the bSQI \cite{JohnsonBehar2015} to applied to select the optimal linear combination of multiple channels. 
Note that this idea can be applied when we have a single channel ta-mECG. We take the publicly available code from \url{http://www.fecgsyn.com} for RLS, LMS, ESN and EKF, and follow the suggested parameters accompanying the code. Denote these methods as $\text{ds-AM}_{\text{RLS}}$, $\text{ds-AM}_{\text{LMS}}$, $\text{ds-AM}_{\text{ESN}}$, and $\text{ds-TS}_{\text{EKF}}$ respectively.

When we only have a single channel ta-mECG, we compare the proposed algorithm with two benchmark template subtraction (TS) algorithms discussed in \cite{Andreotti2016} and the publicly available code provided in \url{http://www.fecgsyn.com}. To run these TS algorithms, we apply Step 2-1 to determine maternal R peaks from the ta-mECG and construct segments associated with all R peaks like \eqref{Equation:ConstructSegments}. 
For the first TS algorithm, we apply PCA to all segments to determine the principle components. For each segment, the mECG is estimated by selecting suitable principal components and applying a back-propagation step on a beat-to-beat basis. See \cite{Andreotti2016} for details. We denote the first TS algorithm as $\text{ds-TS}_{\text{PCA}}$.
For the second TS algorithm, the median of all segments is taken as the template for the mECG. The mECG template is then adapted to each segment using a scalar gain. We denote the second TS algorithm as $\text{ds-TS}_{\text{c}}$

{
Moreover, to appreciate the advantage of the OS step compared with the traditional SVD approach \cite{Oosterom1986, Vander1987, Kanjilal1997}, we consider an algorithm that is the same as our proposed algorithm, except \eqref{CriticalStepOSStep}, where we simply take the top singular vector associated with the largest singular value as the estimated mECG. We denote this algorithm as $\textup{SVD}_{top1}$.
}

\subsection{Evaluation of Morphology Recovery}\label{Section:EvaluationMorphology}

We evaluate the performance of recovering the fECG morphology on the {semi-real} simulated {ta-mECG} signals {of healthy fetus, in both the single channel and two channels cases}. 
{When there is one channel,} the P wave and T peak locations of the simulated fECG and estimated fECG are detected by the algorithm suggested in \cite{bioengineering3040026}.  {When there are two channels, we take the optimal linear combination of two clean simulated fECG's and the estimated fECG to determine P peaks and T peaks. We view the detected P peak and T peak from the clean simulated fECG as the ground truth. The ECG signals for the clean simulated fECG's and the detected P peaks and T peaks are visually inspected to confirm the quality of the true annotation.} 
We follow the same way as evaluating the R peak detection by MAE \cite{Andreotti2016} to evaluate the P peak and T peak detection -- {only those beats with \texttt{TP} R peak detection are considered to evaluate the detected P peak (or T peak).} Then the $\text{F}_1$ and MAE are also reported.

In addition to reporting the performance of detecting P wave and T peak, we report how the fECG morphology is recovered. The {\em normalized mean amplitude error} (NMAE) for R peak morphology recovery is defined as
\begin{equation}
\textup{NMAE} := \frac{1}{n_{\texttt{TP}}} \sum_{i=1}^{n_{\texttt{TP}}} \frac{|\mathbf{z}(t_i) - \tilde{\mathbf{z}}(\tilde{t}_i)|}{|\mathbf{z}(t_i)|}\,,
\end{equation}    
where $\tilde{t}_i$ is the timestamp of the $i$-th \texttt{TP} estimated R peak, $t_i$ is the timestamp of the associated true R peak, and $\mathbf{z}$ and $\tilde{\mathbf{z}}$ represent the whole simulated fECG and its estimation respectively. We also report NMAE for P peaks and T peaks. 
Similarly, to evaluate the performance of recovering PR {(the interval between P peak and R peak)}, QT {(the interval between Q wave and T peak)} and ST {(the interval between S wave and T peak)} intervals, we consider the {\em normalized mean duration error} (NMDE):
\begin{equation}
\textup{NMDE} := \frac{1}{n_{\texttt{TP}}} \sum_{i=1}^{n_{\texttt{TP}}} \frac{|\textup{PR}_i - \widetilde{\textup{PR}}_i|}{\textup{PR}_i}\,,
\end{equation}  
where $\textup{PR}_i$ is the {length of the} $i$-th estimated PR interval and $\widetilde{\textup{PR}}_i$ is the associated {length of the} true PR interval. The same formula hold for QT and ST intervals.
{Again, we follow the same way as evaluating the R peak detection to report NMAE and NMDE -- we only consider those beats with \texttt{TP} R peaks to evaluate the estimated peaks and intervals.} Clearly, NMAE and NMDE jointly represent how much the morphology of the estimated fECG deviates from the ground truth. The smaller the NMAE and NMDE are, the better the performance of the algorithm.
{ To focus on evaluating the performance of fECG morphology recovery, we take the ground-truth fetal R peak annotation into account to determine which decomposed component is the estimation of the fECG.}

\section{Results} \label{Section:Result}
\subsection{Evaluation of R peak location estimation}

The performance of the proposed algorithm for 2 channels tested in \texttt{CinC2013} is summarized in Table \ref{table 1}. The best $\textup{F}_1$ is $93.21 \pm 14.31 \%$, which is achieved in combining channel 1 and channel 4. The associated MAE is $5.44\pm 4.18$ msec.
On the other hand, the $\mathbf{F}_1(1)$ and MAE(1) of all recordings achieves $96.31\pm 10.93\%$ and $4.93 \pm 3.64$ ms. Clearly, for each recording, if we choose among all combinations, the performance is better.
In this table, we also compare the proposed algorithm with SAVER \cite{LiFraschWu2017} and other algorithms, including $\textup{ds-AM}_{\textup{RLS}}$, $\textup{ds-AM}_{\textup{LMS}}$, $\textup{ds-AM}_{\textup{ESN}}$, $\textup{ds-TS}_{\textup{EKF}}$, $\textup{ds-TS}_{\textup{PCA}}$, $\textup{BSS}_{\textup{ICA}}$, {$\textup{SVD}_{top1}$}, and we see that the proposed algorithm outperforms all other algorithms, including SAVER. {Specifically, if we consider the $\textup{SVD}_{top1}$ algorithm, that is, we replace the OS step by the top singular vector, the overall performance drops. This indicates the benefit of introducing the OS.}

When there are more channels, the proposed algorithm should lead to better results. To confirm this fact, we report the results when we have three channels. The result is summarized in Table \ref{table 3}. The best $\textup{F}_1$ is $93.91 \pm 14.83 \%$, which is achieved when we combine channels 1, 2 and 4. The associated MAE is $5.64\pm 5.11$ msec. We see an improvement over the best two-channel combination, channels 1 and 4, when we include channel 2.
On the other hand, the $\mathbf{F}_1(1)$ and MAE(1) of all recordings achieve $95.32\pm 13.75\%$ and $5.41 \pm 5.22$ ms. Since there are three channels, we also compare with available blind source separation algorithms. Again, we run the publicly available code provided by \cite{Andreotti2016}, {and we conclude that} our result is consistently better, including $\mathbf{F}_1(1)$ and MAE(1).

The proposed algorithm is also applicable when we have only one channel. The result is summarized in Table \ref{table 2}. The best $\textup{F}_1$ is $75.63 \pm 30.96 \%$, which is achieved when channel 2 is considered. The associated MAE is $8.97\pm 7.77$ msec.
On the other hand, the $\mathbf{F}_1(1)$ and MAE(1) of all recordings achieves $88.05\pm 23.08\%$ and $6.08 \pm 5.41$ ms. Overall, we can see a performance enhancement after introducing OS when compared with the algorithm shown in \cite{SuWu2017}. We also compare the result with the state-of-the-art template subtraction algorithm. For a fair comparison, we run the publicly available template subtraction code \cite{Andreotti2016}\footnote{\url{http://www.fecgsyn.com}}. Clearly, our result is consistently better over all channels, as well as $\mathbf{F}_1(1)$ and MAE(1). {Again, the performance of the $\textup{SVD}_{top1}$ algorithm, is worse, which again indicates the benefit of the OS step.}

{To further demonstrate the strength of the proposed algorithm, we consider a more stringent evaluation criteria. In the above reported results, the $F_1$ defined in \eqref{Definition F1} depends on a matching window of $50$ ms \cite{AndreottiRiedl2014}. Now we report the $F_1$ based on two smaller matching windows, one of $25$ ms and one of $10$ ms. The result is shown in Table \ref{Table:Different Windows}. When the matching window is $25$ ms ($10$ ms resp.), the best $\textup{F}_1$ is $89.95 \pm 16.98 \%$ ($84.15\pm 20.65$ resp.), which is achieved in combining channel 1 and channel 4. The associated MAE is $4.04\pm 2.60$ ($3.36\pm 2.15$ resp.) msec.
On the other hand, the $\mathbf{F}_1(1)$ and MAE(1) of all recordings achieves $94.05\pm 13.16\%$ and $3.92 \pm 2.50$ msec ($90.67\pm 14.83\%$ and $3.21 \pm 1.99$ msec resp.). 
Clearly, when the matching window is smaller, the $F_1$ decreases but the MAE also decreases. This result indicates that if we choose the matching window to be $25$ msec, the performance is still better than the state-of-the-art algorithm SAVER \cite{LiFraschWu2017}. Even when the matching window is $10$ msec, the $F_1$ result is slightly worse but comparable.}

\subsection{Demonstration of fECG morphology recovery}\label{Section:Demonstration of fECG recovery}

\begin{figure}[ht]
\center
\includegraphics[width=.85\textwidth]{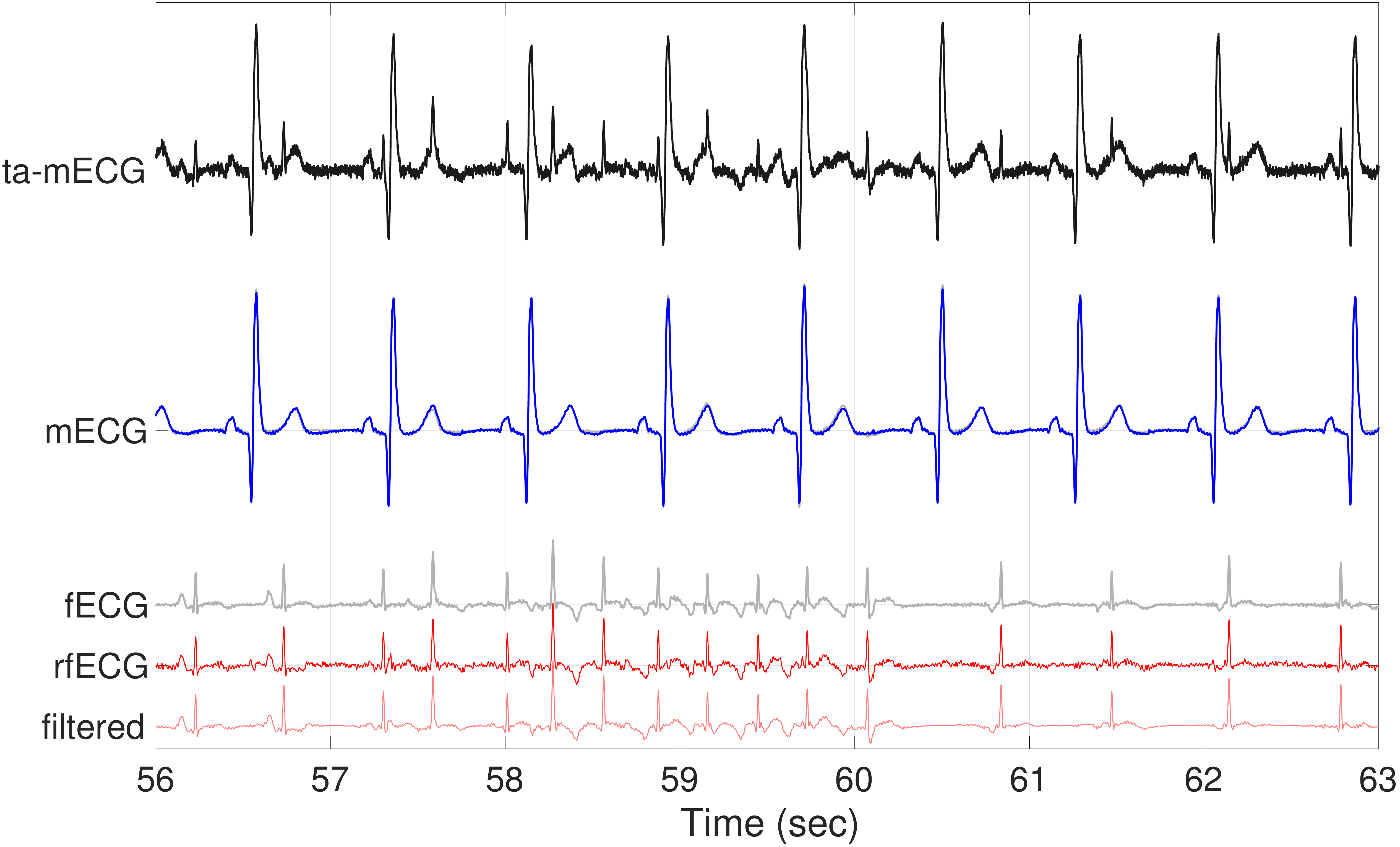}
\includegraphics[width=.85\textwidth]{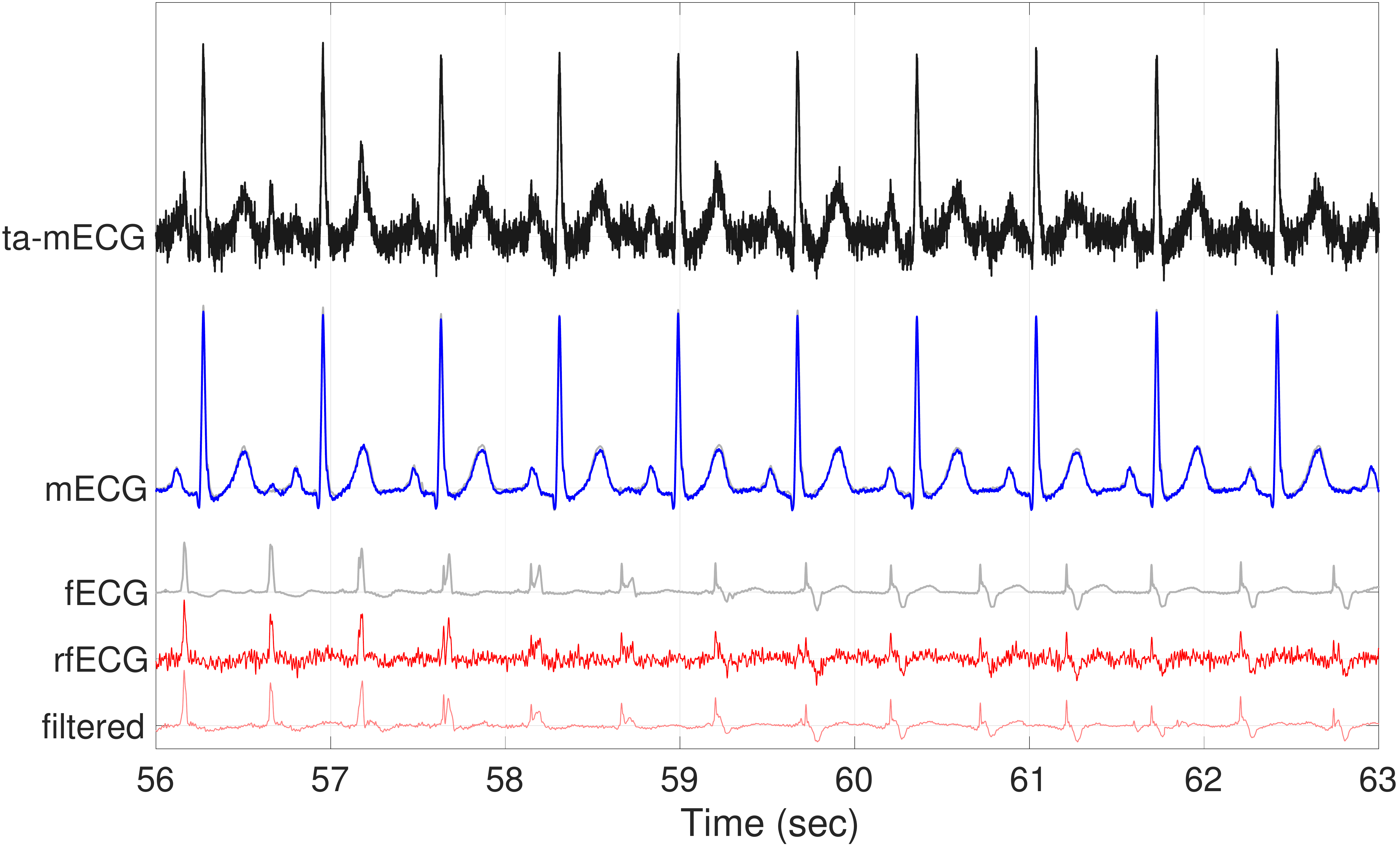}
\caption{An illustration of fetal ECG (fECG) recovery from the single channel simulated trans-abdominal maternal ECG (mECG) recordings for two arrhythmic fetuses. 
Both signals are of length 57 seconds. In the top figure, the signal-to-noise ratio (SNR) is $20$ dB and the simulated fECG amplitude is $1/4$ of the mECG. In the bottom figure, the SNR is $10$ dB and the simulated fECG amplitude is $1/6$ of the mECG. The black tracking is the simulated ta-mECG, the gray trackings in the middle and bottom are the simulated mECG and fECG, the blue and red trackings are the estimated mECG and rough fECG (rfECG), and the light red tracking in the bottom is the denoised fECG by the optimal shrinkage.
We can easily identify the landmarks P, Q and T in the sinus beats and those arrhythmic beats. It is not surprising that the larger the fetal amplitude and SNR, the cleaner the estimated fECG is.}
\label{Figure:Recovery2}
\end{figure}

See Figure \ref{Figure:Recovery2} for an illustration of the reconstructed fECG from two single channel simulated ta-mECGs. The SNR is 20dB with the amplitude of the simulated fECG $1/4$ of the simulated mECG in the first case, and the SNR is 10dB with the amplitude of the simulated fECG $1/6$ of the simulated mECG in the second case. Overall, while the morphology is slightly deviated from the simulated one, the main landmarks are easy to identify in the reconstructed signal. We see that even if the maternal and fetal QRS complexes overlap, we can well recover the fetal ECG morphology.  
However, in some cases, particularly when the fECG amplitude is small and the SNR is low, the morphology may have a large deviation. We can further enhance the rough fECG quality by applying the OS. The results are shown as light red signals in the bottom of Figure \ref{Figure:Recovery2}. We see that the OS does help improve the quality of the rough fECG for the visual inspection purpose. %

\begin{figure}[ht]
\center
\includegraphics[width=.85\textwidth]{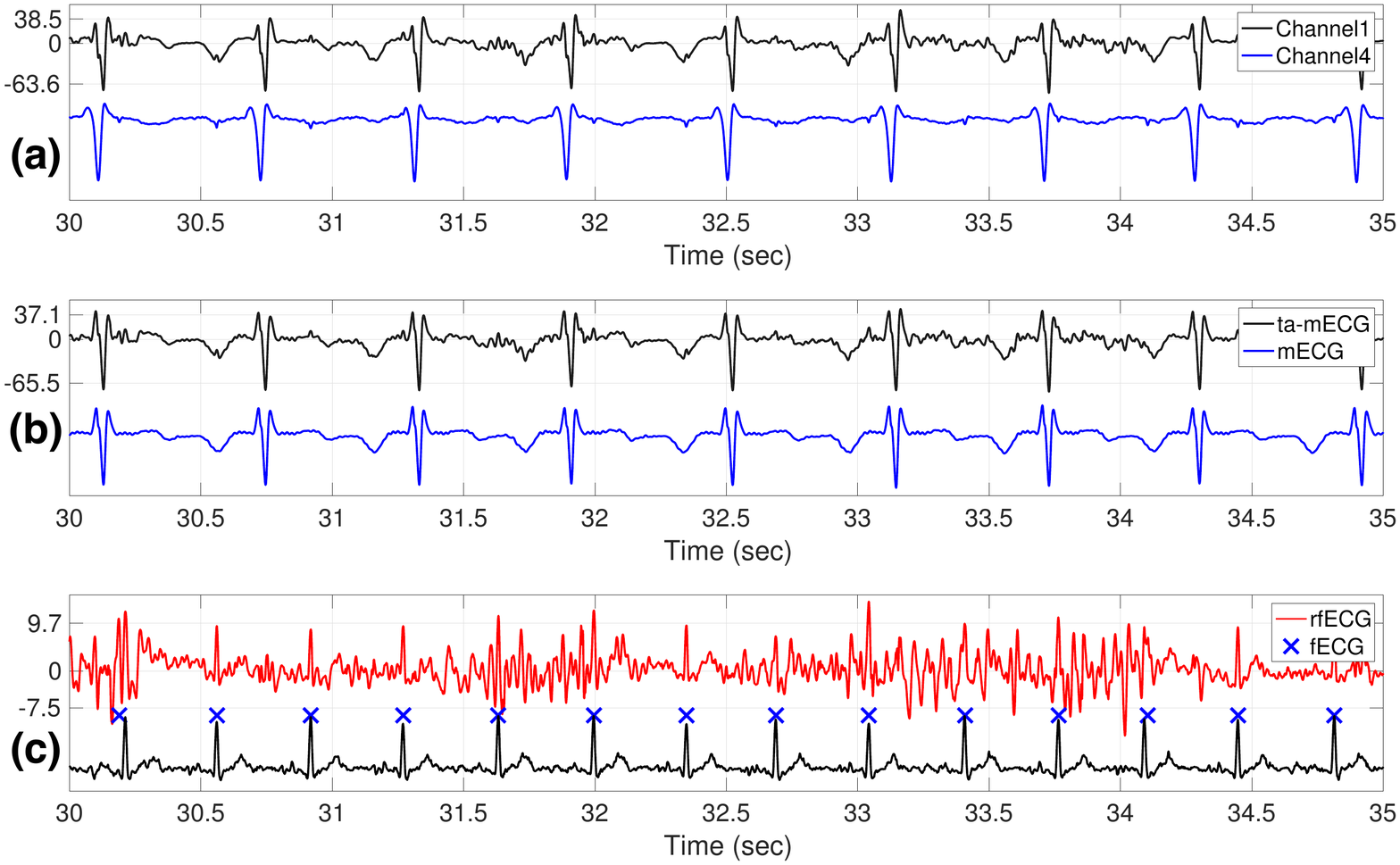}
\includegraphics[width=.85\textwidth]{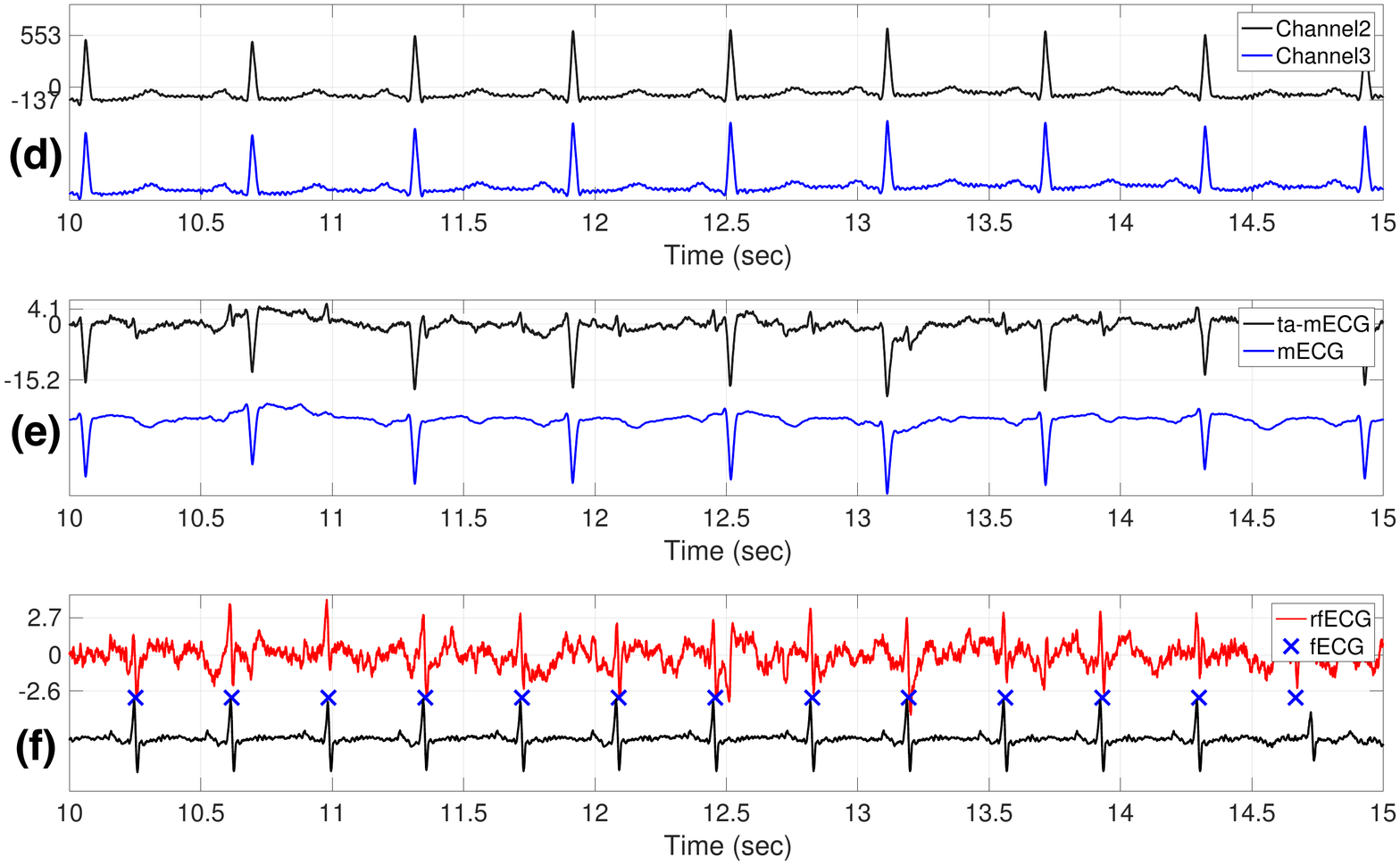}
\caption{An illustration of fetal ECG (fECG) recovery from the trans-abdominal maternal ECG (mECG) recordings from the \texttt{CinC2013} database. 
The subplots (a), (b) and (c) come from a linear combination of Channels 1 and 4 of the 6-th recording, and the subplots (d), (e) and (f) come from a linear combination of Channels 2 and 3 of the 59-th recording. In (b) and (e), the black tracking is the linearly combined ta-mECG, the blue tracking is the estimated mECG; in (c) and (f), the red tracking is the rough fECG (rfECG), the black tracking is the estimated fECG (fECG) depending on the optimal shrinkage, the blue crosses are the provided labels by the experts.
We can easily identify the landmarks P, Q and T in the sinus beats, even if the noise level is relatively high for the fECG like in the 6-th recording.}
\label{Figure:RecoveryCinC2013}
\end{figure}

{The decomposition result on the real databases, \texttt{CinC2013} and \texttt{nifeadb}, are shown in Figures \ref{Figure:RecoveryCinC2013} and \ref{Figure:RecoveryNIFEADB}. For the normal fetus from \texttt{CinC2013}, the landmarks P, Q and T can be easily identified, even if the noise level is relatively high for the fECG in the 6-th recording. For the 59-th recording, it is hard to see any fECG in Channel 2 and Channel 3, but after suppressing the mECG by the linear combination idea, we are able to see the fECG in the linearly combined ta-mECG, and hence the extraction of the fECG. In this case, we can see clear P waves in the final recovered fECG, while the T wave is relatively weak.

For the arrhythmic fetus in \texttt{nifeadb}, the results are demonstrated in Figure \ref{Figure:RecoveryNIFEADB}. We show the same cases, ARR 3, ARR 4 and ARR 11, demonstrated in \cite[Figures 3, 4 and 5]{behar2019noninvasive}. In ARR 3, we can easily identify the trigeminy pattern, and the P-waves are visible in most beats, indicated by the blue arrows, allowing perinatal cardiologists to characterize the premature atrial contraction (PACs). Compared with \cite[Figure 3]{behar2019noninvasive}, the reconstructed P-waves is more viable from beat to beat. This is expected since we only use two channels. In ARR 4, the blocked P-wave can be easily visualized. Note that compared with the extracted fECG from the multichannel ta-mECG signals shown in \cite[Figure 4]{behar2019noninvasive}, the blocked normal P-wave is less clear in our recovery. In ARR 11, the atrial tachycardia with the 2:1 AV conduction can be easily identified. 
Overall, a side-by-side comparison of the extracted fECG's by our algorithm with those extracted by the algorithm taking multiple channels into account shown in \cite[Figures 3, 4 and 5]{behar2019noninvasive}, we see that the overall quality of our result is not superior but comparable. Particularly, the proposed algorithm could reconstruct most important structure prenatal cardiologists have interest. This result shows the potential of the proposed algorithm for the arrhythmia diagnosis. 
}

\begin{figure*}[ht]
\center
\includegraphics[width=.8\textwidth]{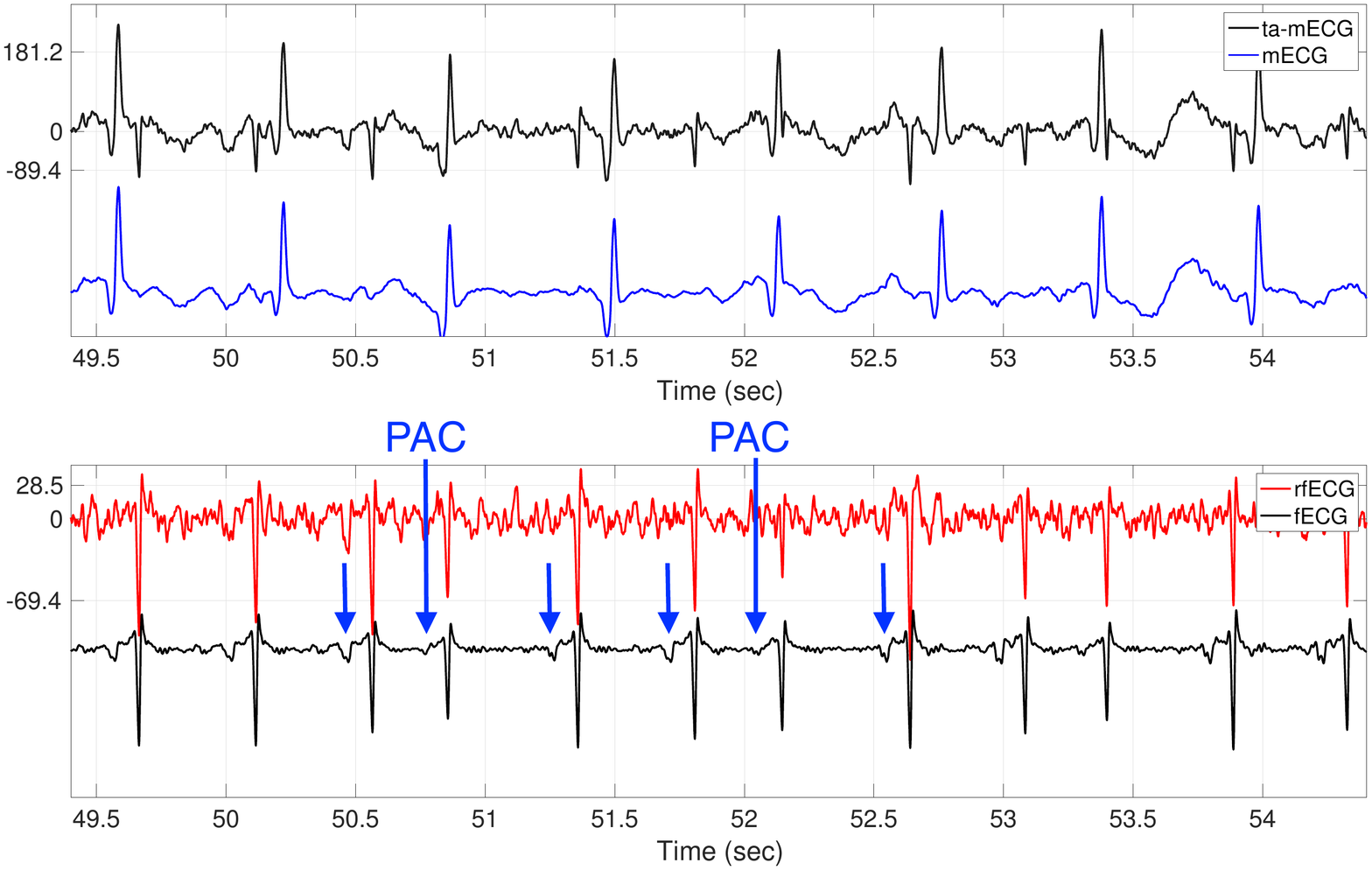}
\includegraphics[width=.8\textwidth]{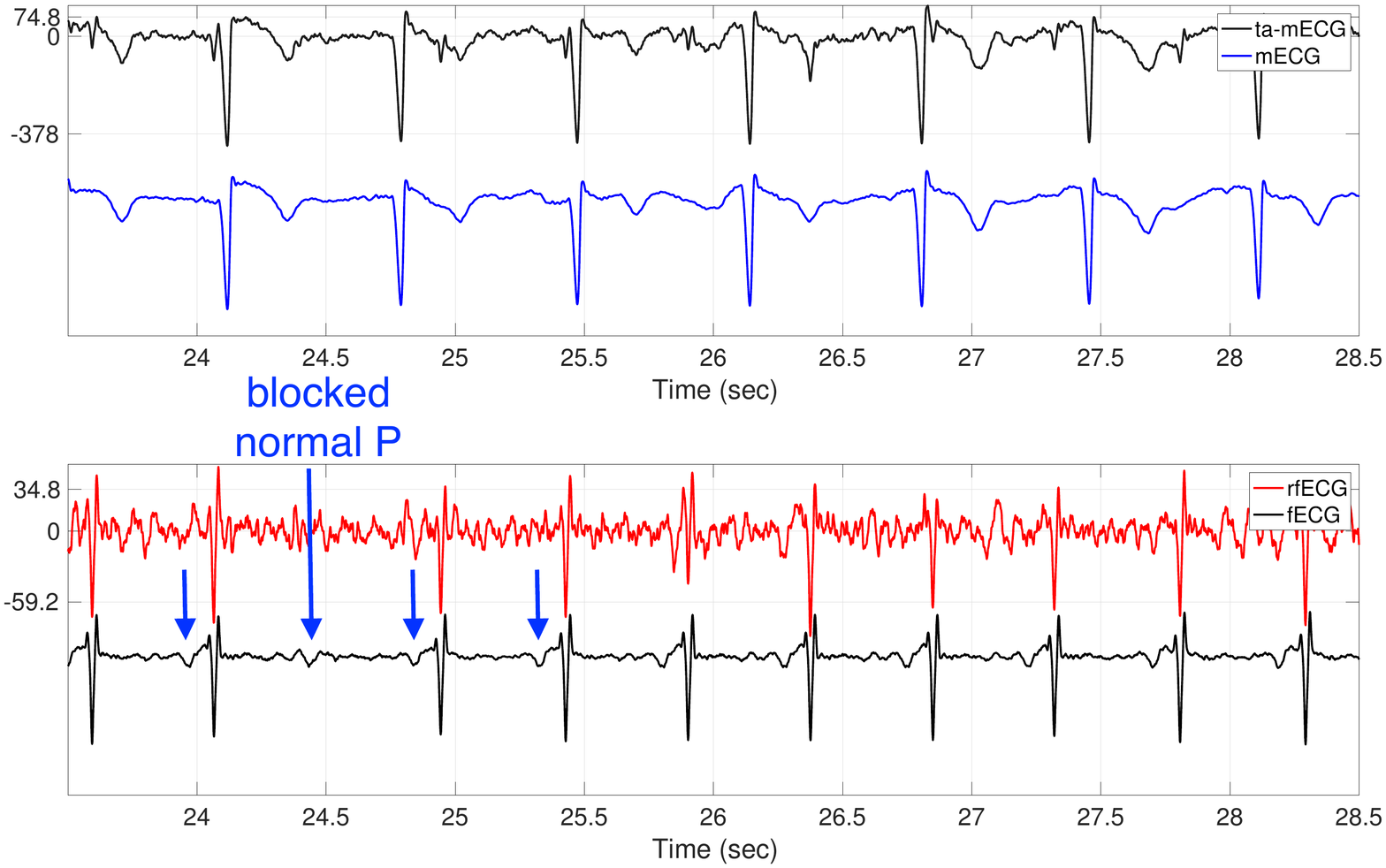}
\includegraphics[width=.8\textwidth]{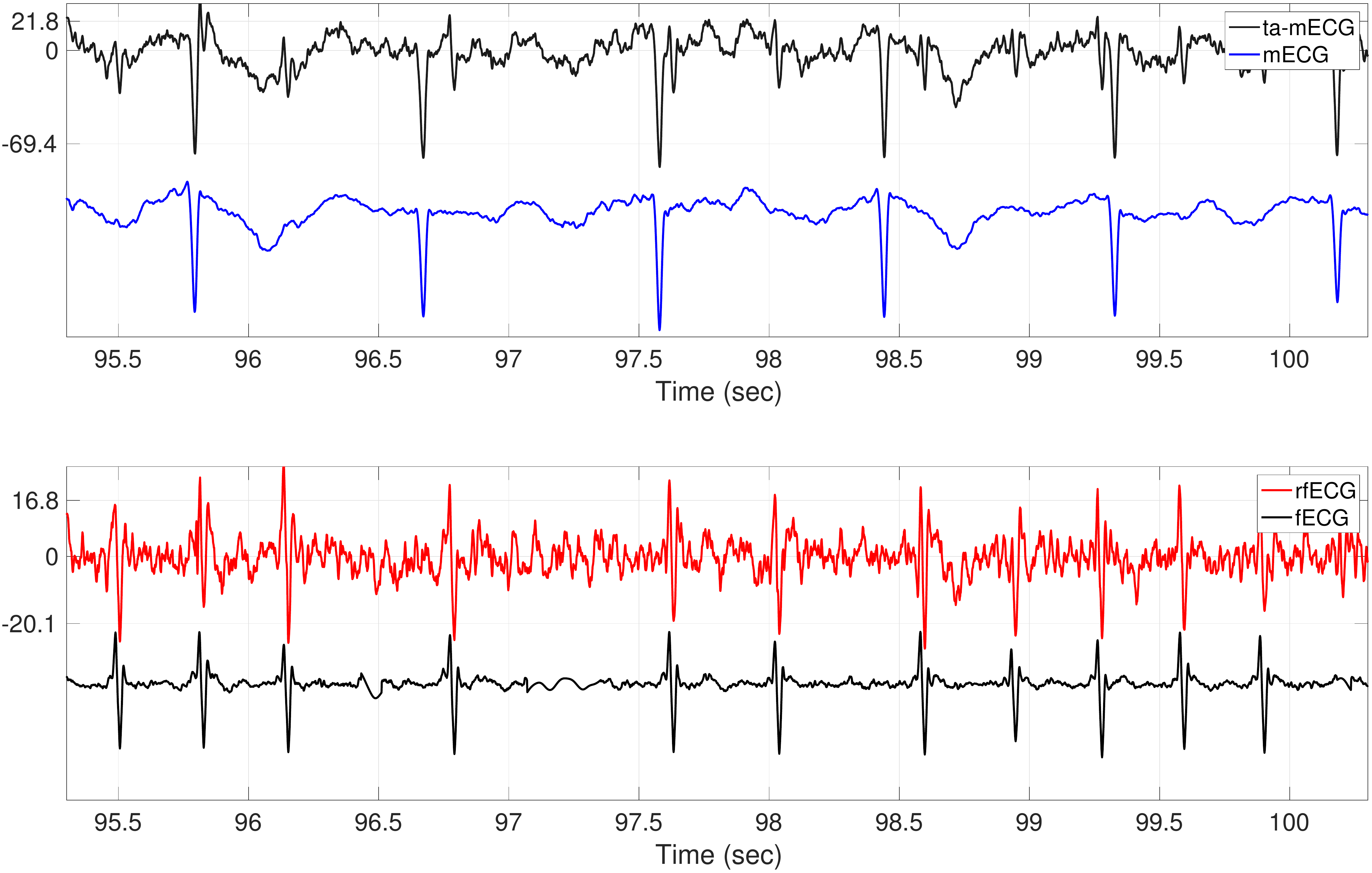}
\end{figure*}

\begin{figure}
\caption{An illustration of fetal ECG (fECG) recovery from the trans-abdominal maternal ECG (mECG) recordings from the \texttt{nifeadb} database. The optional nonlocal median filter step is applied to enhance the fECG.
The subplots (a) and (b) come from a linear combination of Channels 3 and 4 of ARR 3, the subplots (c) and (d) come from a linear combination of Channels 1 and 3 of ARR 4, and the subplots (e) and (f) come from a linear combination of Channels 2 and 4 of ARR 11. In (a), (c) and (e), the black tracking is the linearly combined ta-mECG, the blue tracking is the estimated mECG; in (b), (d) and (f), the red tracking is the rough fECG (rfECG) and the black tracking is the estimated fECG (fECG) depending on the optimal shrinkage.
In ARR 3, we can easily identify the trigeminy pattern, and the P-waves are visible allowing to characterize the premature atrial contraction (PACs). In ARR 4, the blocked P-wave can be visualized. In ARR 11, we see the atrial tachycardia with 2:1 AV conduction.}
\label{Figure:RecoveryNIFEADB}
\end{figure}

\subsection{Quantification of fECG morphology recovery}\label{Section:Quantification of fECG recovery}

\begin{figure}[ht]
\center
\includegraphics[width=.95\textwidth]{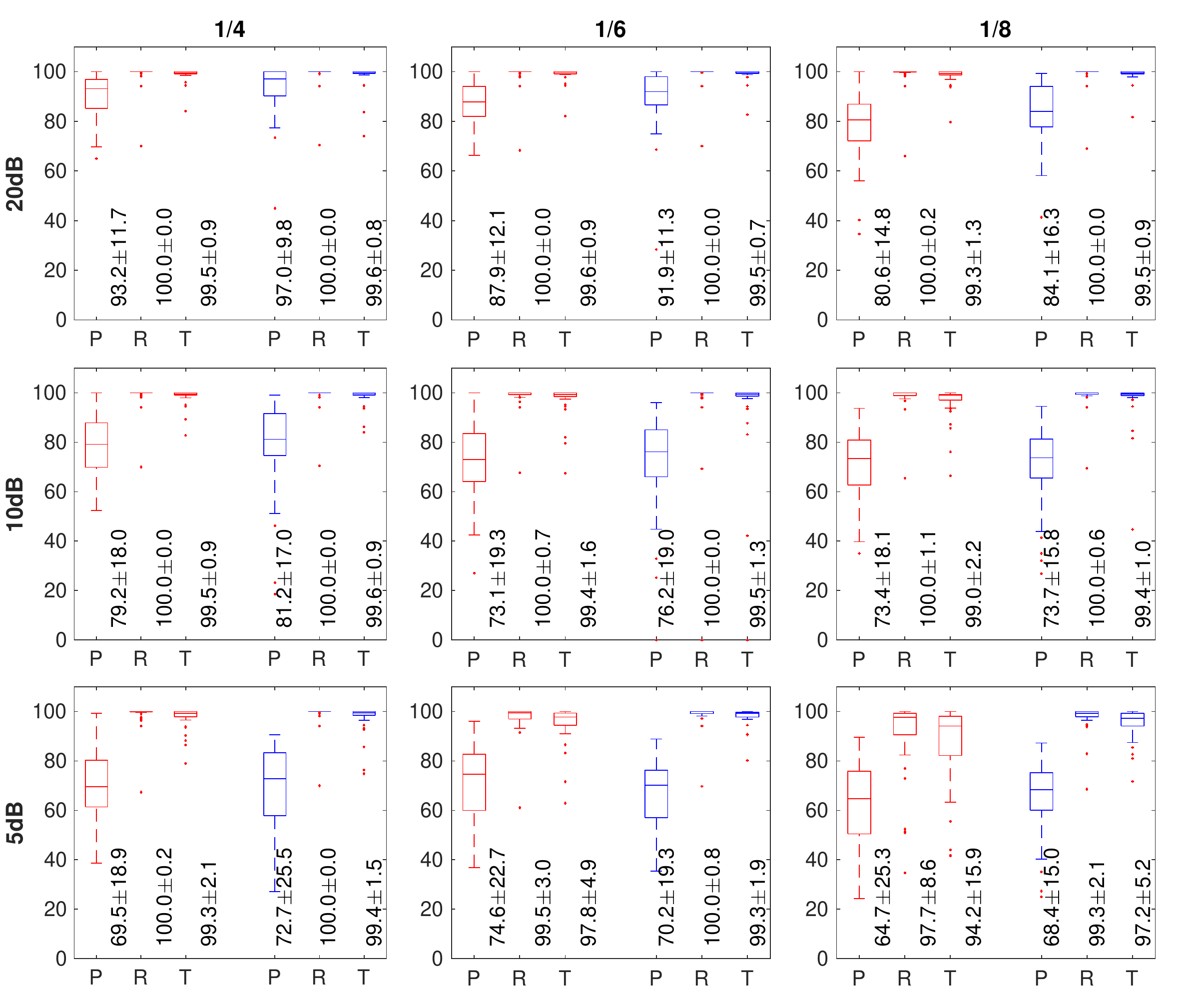}
\caption{The boxplot of $\textup{F}_1$ of recovering P wave, R peak and T peak by the proposed algorithm. The rows indicate the ratio of the fetal ECG (fECG) amplitude and the maternal ECG (mECG) amplitude, {ranging from $1/4$, $1/6$ and $1/8$ (left to right columns), and the columns indicate the signal-to-noise ratio (SNR), ranging from $20$ dB, $10$ dB and $5$ dB (top to bottom rows). The red bars on the left hand side are results from the single channel simulation, and the blue bars on the right hand side are results from the two channels simulation.}  
The circle indicates the median, the thick bar indicates the interquartile range, the dots indicates the outliers, and the thin bar indicates the range except the outliers. As expected, when the ratio of the fECG and the mECG is high and the SNR is high, the $\textup{F}_1$ is high.}
\label{Figure:BoxPlot_f1}
\end{figure}

\begin{figure}[ht]
\center
\includegraphics[width=.95\textwidth]{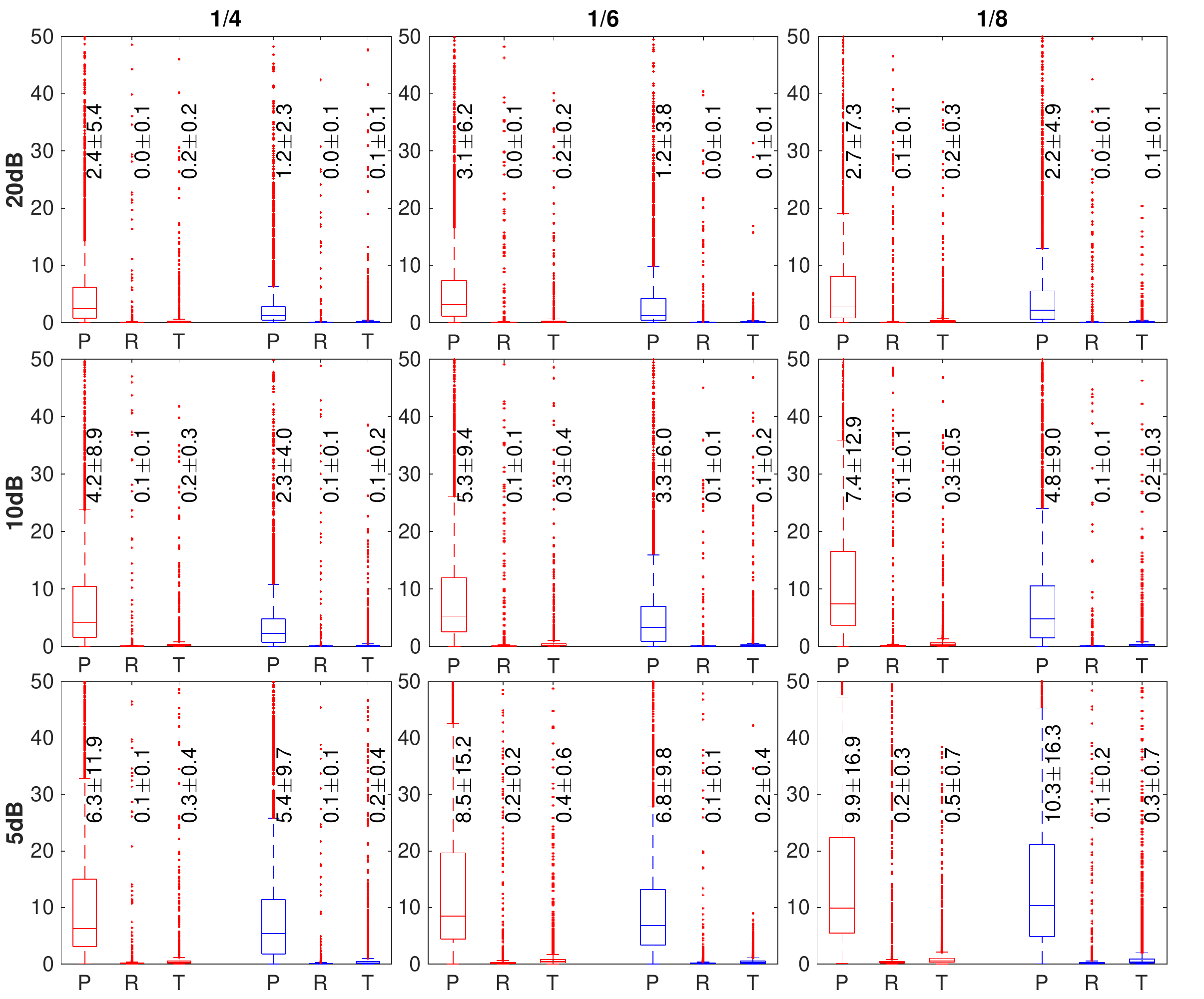}
\caption{The boxplot of normalized mean amplitude error (NMAE) of recovering P wave, R peak and T peak by the proposed algorithm. The rows indicate the ratio of the fetal ECG (fECG) amplitude and the maternal ECG (mECG) amplitude, {ranging from $1/4$, $1/6$ and $1/8$ (left to right columns), and the columns indicate the signal-to-noise ratio (SNR), ranging from $20$ dB, $10$ dB and $5$ dB (top to bottom rows). The red bars on the left hand side are results from the single channel simulation, and the blue bars on the right hand side are results from the two channels simulation.} 
The circle indicates the median, the thick bar indicates the interquartile range, the dots indicates the outliers, and the thin bar indicates the range except the outliers. 
As expected, when the ratio of the fECG and the mECG is high and the SNR is high, the NMAE is close to $0$. }
\label{Figure:BoxPlot_ar}
\end{figure}

\begin{figure}[ht]
\center
\includegraphics[width=.95\textwidth]{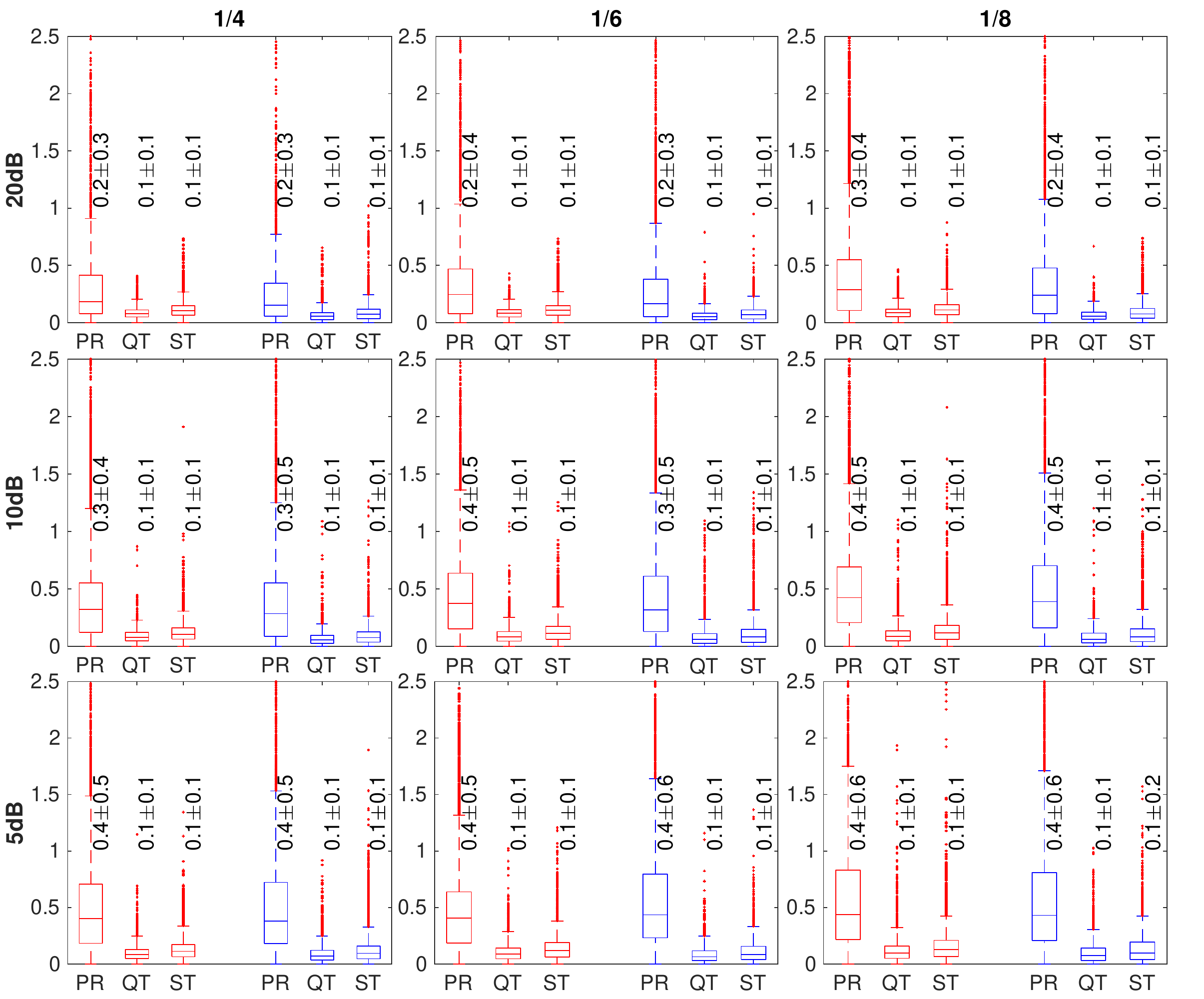}
\caption{The boxplot of normalized mean duration error (NMDE) of recovering PR interval, QT interval and ST segment by the proposed algorithm. The rows indicate the ratio of the fetal ECG (fECG) amplitude and the maternal ECG (mECG) amplitude, {ranging from $1/4$, $1/6$ and $1/8$ (left to right columns), and the columns indicate the signal-to-noise ratio (SNR), ranging from $20$ dB, $10$ dB and $5$ dB (top to bottom rows). The red bars on the left hand side are results from the single channel simulation, and the blue bars on the right hand side are results from the two channels simulation.}
The circle indicates the median, the thick bar indicates the interquartile range, the dots indicates the outliers, and the thin bar indicates the range except the outliers.  As expected, when the ratio of the fECG and the mECG is high and the SNR is high, the NMDE is close to $0$.}
\label{Figure:BoxPlot_dr}
\end{figure}

{To further quantify the results, $\textup{F}_1$, NMAE and NMDE evaluated from the estimated fECG are shown in} Figures \ref{Figure:BoxPlot_f1}, \ref{Figure:BoxPlot_ar} and \ref{Figure:BoxPlot_dr} under different ratios of fECG and mECG and SNR's {for a direct comparison}. Here we follow the report scheme suggested in \cite{Andreotti2016}. In the boxplot, the circle indicates the median, thick line indicates the interquartile range, the dots indicate the outliers, while the thin line indicates the range of the data without the outliers. Here, the outliers are defined as those values that are outside $[Q_1-q(Q_3-Q_1),Q_3+q(Q_3-Q_1)]$, where $Q_1$ and $Q_3$ are the $25$-th and $75$-th percentiles of the sample data, and $q$ is chosen to be $1.5$. 
As expected, the higher the ratio of the fECG and the mECG, and the higher the SNR, the higher the $\textup{F}_1$ and the lower the {NMAE and NMDE. Compared with one channel, the morphology recovery performance is in general better when we have two channels. Since a P wave in general has a smaller amplitude compared with R and T waves, it is not surprising to see that all quantities involving a P wave are less accurate with more outliers and more vulnerable to a small SNR and a small ratio}.


\section{Discussion and Conclusion} \label{Section:Discussion}

We provide a novel algorithm to recover fECG morphology from few ta-mECG channels. Compared with the traditional algorithms, the main novelty is introducing a new metric to compare cardiac activities based on the OS theory. 
The algorithm is supported by solid mathematical foundation, and gives a convincing fetal R peak detection result on the publicly available databases compared with those reported in \cite{SuWu2017,LiFraschWu2017,shnitzer2018recovering}, and the fECG morphology recovery result is confirmed on a semi-real simulated database. 
For the fECG morphology recovery, an important advantage of the proposed algorithm is the ability to recover fECG even when the fetal QRS overlaps a maternal QRS.

\subsection{Clinical application and significance}

Compared with the state-of-the-art result when we have only few channels \cite{SuWu2017,LiFraschWu2017}, the proposed algorithm leads to a more accurate result. This more accurate fetal R peak detection allows better fetal heart rate analysis. Specifically, when two channels are available, the MAE of the best combination is 5.45 $\pm$ 4.65 msec. This is equivalent to extracting a R peak to R peak interval from an ECG signal sampled at about 150Hz. With this improvement compared to the state-of-the-art algorithm SAVER when 2 channels are available, we are closer to the recommendation for the ECG sampling rate proposed in \cite{TaskForce:1996} for the adult HRV analysis. 

To the best of our knowledge, this is the first work showing the possibility to recover the fECG morphology with high accuracy from few ta-mECG channel signals, even when the fetus has an arrhythmia. 
It has the potential to help fetal ST analysis monitoring that detects and alerts fetal hypoxia, the diagnosis of long QT syndrome that may not result in obvious arrhythmias on echocardiography, and elucidate the origin and mechanisms of tachyarrhythmias and other conditions. 
Future work will need to focus on real-time clinical data to assess applicability of the method for patient clinical use.
Since only few channels are needed, {the algorithm has the potential to help long term monitoring of fetal status. This cannot be achieved by periodic fetal echocardiography.}

\subsection{Comparison with other algorithms}

To the best of our knowledge, there are limited algorithms available to analyze fECG when two or three ta-mECG channels are available, except SAVER \cite{LiFraschWu2017} and the diffusion-based approach \cite{shnitzer2018recovering}. 
In general, the proposed algorithm falls in the category of the TS method. 
To have a more systematic comparison and understanding of the relationship, note that the proposed algorithm contains three main ingredients:
\begin{enumerate}
\item de-shape short time Fourier transform approach to detect maternal R peaks;
\item linear combination of two or three channels, and the bSQI selection of the best channel;
\item OS approach to determine neighbors for the nonlocal median purpose.
\end{enumerate}
The first ingredient has been extensively discussed in the previous paper \cite{SuWu2017} and we refer interested readers there. This ingredient is universal and can be combined with other TS methods.

The second ingredient, the linear combination based on the dipole current model, is first proposed in \cite{LiFraschWu2017} {to handle the case when there are only two ta-mECG channels}. In \cite{LiFraschWu2017}, the optimal linear combination is determined by {combining the lag map and diffusion map \cite[(2.2)]{LiFraschWu2017}}. However, we found that a more straightforward bSQI helps determine the optimal linear combination more efficiently -- with the bSQI, $\text{ds-TS}_{\text{PCA}}$ outperforms SAVER when we have two channels. See Table \ref{table 1} for details. Thus, in this paper the bSQI is proposed to replace the {combination of lag map and diffusion map} approach. We mention that this linear combination idea has the potential to be combined with algorithms, like ICA or AF-based algorithms. We will explore this possibility in the future work.

The third ingredient is the main ingredient that can be directly compared with the traditional TS method and its several variations. In the traditional TS method and its variations, the mean, principal components, or singular vectors, of {\em consecutive} cardiac activities in the ta-mECG are considered to construct the template of the cardiac activity. 
These traditional approaches do not take into account the fact that the QRST complex morphology (both maternal and fetal) is time-varying \cite{MalikFarbom2002}. In the proposed algorithm, we instead {model} similar QRST complexes located at different time by {a low rank model so that the} OS {can be applied} to recover the ECG signal. 
%

In Tables \ref{table 1} and \ref{table 3}, we see a consistent low performance of the BSS approach like ICA. This is not surprising -- it is well known that the performance of ICA might be limited when there are only 2 or 3 channels, since usually we need more than 4 channels to have a reasonable result \cite{Andreotti2016}.

\subsection{Limitation and future work}

The main weak point of this work is the lack of a large scale clinical database with clinical outcome. Specifically, for this study we had to use an existing publicly available benchmark database to demonstrate the potential of the proposed algorithm. While it allows us to compare the {fetal R peaks estimation} result with other algorithms, it does not contain enough information for us to make further conclusion {for the fECG morphology reconstruction}. For example, although we have a reasonable reconstruction of PR, ST and QT intervals in the semi-real simulation database {with a quantitative measurement in Section \ref{Section:EvaluationMorphology}}, we do not have the PR, ST or QT interval information commonly used in clinics for the \texttt{CinC2013} {and \texttt{nifeadb} databases}. 
Moreover, while we show several decomposed fECG's for data in \texttt{CinC2013} and \texttt{nifeadb} databases in Section \ref{Section:Demonstration of fECG recovery}, we do not have a systematic evaluation of the clinical relevance of the reconstructed fECG, like those results shown in \cite{CliffordSameni2011,behar2019noninvasive}.
So we cannot conclude how accurate the estimated PR, {ST} or QT interval is and what clinical information we can provide {from the reconstructed fECG morphology}. Thus, collecting a large scale and prospective dataset with necessary clinical information and {carrying out a clinical application with our clinical team are} urgently needed. {Moreover, although only two channels are considered, our reconstruction results are in some sense comparable with the results shown in \cite[Figures 3, 4 and 5]{behar2019noninvasive}; specifically, our methods could reconstruct the critical structures. A more systematic study of the \texttt{nifeadb} database, as well as the above-mentioned limitations, will be reported in our future work.}

In the simulation, we take the semi-real ta-mECG to evaluate the performance of recovering the fECG morphology by the proposed algorithm. While this semi-real approach is closer to the real world scenario compared with the model-based simulation proposed in \cite{Andreotti2016}, it has its own limitation; particularly, note that the projection directions of simulated mECG's for two channels are fixed, the simulated fECG's comes from fixed channels, and the respiratory effect on the cardiac axis is not corrected, the simulated signal. Each of these facts may deteriorate the rich physiological dynamics inherited in the real-world ta-mECG. We may combine the model-based simulation \cite{Andreotti2016} to develop a more delicate simulation in the future work.

{From the clinical perspective,} another unsolved challenge is interpreting the reconstructed fECG signal. Due to the anatomic variation among pregnant women, like uterus position and shape, as well as the fetal size and presentation, even if we could standardize the lead system on the mother's abdomen, the fECG waveform morphology still varies from subject to subject. Note that the PR interval and QT interval information might be deformed due to this variation, and hence its clinical applicability might be degraded. This situation is more challenging when the gestational age is small. Developing an adaptive algorithm to recover the fetal vectocardiogram for the purpose of establishing an {\em intrinsic fECG lead system} is the next challenge we face.
{From the clinical treatment perspective, we would like to monitor the fECG for medication effects -- either medications given to treat the mother or medications given to the mother to treat the fetal arrhythmia. We could monitor fetal heart rates to see if we control the arrhythmia but to the best of our knowledge, so far we have no ability to monitor the effects of medications on fetal conduction.}

From the algorithm design perspective, theoretically, we may further take the ECG morphology knowledge into account to design a more robust and accurate OS. This direction involves more theoretical work about ``wave-shape manifold'' analysis \cite{lin2019wave}. Due to the nonlocal nature of the proposed algorithm, it can be easily turned into a real-time one for fetal monitoring. {Also, when there are multiple channels available, we could modify the proposed algorithm to take all channels into account to achieve a better fECG morphology recovery. 
Another challenge is an automatic determination of the final fECG estimate from the decomposed two components. This is a commonly encountered channel selection challenge in the BSS algorithm, and usually we need extra information to solve the challenge. In the proposed algorithm, we use a simple approach -- select the decomposed component with the smaller median R-R interval as a fECG estimate, and hence the other one as a mECG estimate. However, when the mother has a higher heart rate, particularly when the fetus has a bradycardia, this method would fail. Note that on the bedside, we can count on physicians' experience to make a decision, but for the real-time diagnosis purpose, we need to find a better criteria to resolve this challenge. }
In this paper, we follow conventional approach and apply the lowpass filter with the cutoff 100Hz in Step 1. While the proposed algorithm provides an encouraging result, it is not clear how much the lowpass filter impacts the recovered fECG morphology. A more detailed spectral content study of the fECG, following \cite{Abboud1989}, is needed to better understand how to design the optimal lowpass filter to recover the fECG morphology.
We leave these challenges to future work.

\section*{Reference}

\bibliographystyle{IEEEtran}
\bibliography{refs,fECGCepstrum}

\clearpage

\begin{table} 
\scriptsize\centering
\caption{The comparison of $\textup{F}_1$ and MAE of different algorithms applied to \texttt{CinC2013} when only two channels are considered. The subject a54 is removed from the dataset. All data are presented as mean$\pm$standard deviation.}
{
\begin{tabular}{|c|c|c|c|c|}
\multicolumn{5}{c}{\vspace{-3pt}}\\
\cline{2-5}
\multicolumn{1}{c|}{} & channels & SAVER \cite{LiFraschWu2017} & {Proposed} & $\textup{ds-AM}_{\textup{RLS}}$ \\
\hline
\multirow{6}{*}{$\textup{F}_1$ (\%)} & 1+2 &81.69$\pm$ 25.82  & {\bf 86.62 $\pm$ 23.34} & 61.05 $\pm$ 33.03\\
& 1+3 & 82.93 $\pm$ 26.28 &  {\bf 89.72 $\pm$ 21.76} & 57.18 $\pm$ 31.88\\
& 1+4 & 87.93 $\pm$ 22.64 & {\bf 93.21 $\pm$ 14.31} & 63.93 $\pm$ 32.42\\
& 2+3 & 74.40 $\pm$ 30.63 &  79.25 $\pm$ 31.75 & 58.01 $\pm$ 33.85\\
& 2+4 & 81.50 $\pm$26.64 & {\bf 87.32$\pm$ 24.44} & 62.28 $\pm$ 32.16\\
& 3+4 & 79.83 $\pm$ 28.49 & {\bf 82.60 $\pm$ 28.03} & 56.23 $\pm$ 32.90\\
\hline
\multicolumn{2}{|c|}{$\textup{F}_1(1)$ (\%)} &92.99$\pm$16.00 & {\bf 96.31$\pm$ 10.93} & 73.60 $\pm$ 27.73\\
\hline
\multicolumn{2}{|c|}{$\textup{F}_1(0.5)$ (\%)} &85.43$\pm$ 22.42 & {\bf 87.94 $\pm$ 21.30} & 62.32 $\pm$ 30.28\\
\hline
\hline
\multirow{6}{*}{MAE (ms)} & 1+2 & 7.72$\pm$ 7.03 & {\bf 6.86 $\pm$ 6.55} & 11.02 $\pm$ 9.30\\
& 1+3 & 7.83 $\pm$ 7.45 & {\bf 6.43 $\pm$ 6.37} &  11.60 $\pm$ 9.21\\
& 1+4 & 6.21 $\pm$ 6.03 & {\bf 5.44 $\pm$ 4.18} &  9.77 $\pm$ 8.75\\
& 2+3 & 9.44 $\pm$ 6.88 & {\bf 8.61 $\pm$ 7.51} &  12.53 $\pm$ 9.66\\
& 2+4 & 7.93 $\pm$ 6.62 & {\bf 7.11$\pm$ 6.77} &  10.59 $\pm$ 8.39\\
& 3+4 & 7.85 $\pm$ 6.85 & {\bf 7.27 $\pm$ 7.29} &  12.51 $\pm$ 9.21\\
\hline
\multicolumn{2}{|c|}{MAE(1) (ms)} & 5.38$\pm$ 4.52& {\bf 4.93$\pm$3.64} & 8.29 $\pm$ 7.68\\
\hline
\multicolumn{2}{|c|}{MAE(0.5) (ms)}& 6.54 $\pm$ 4.92& {\bf 6.37 $\pm$ 5.71} & 9.69 $\pm$ 7.46 \\
\hline

\multicolumn{5}{c}{\vspace{-2pt}}\\
\cline{2-5}
\multicolumn{1}{c|}{} & channels & $\textup{ds-AM}_{\textup{LMS}}$ & $\textup{ds-AM}_{\textup{ESN}}$ & $\textup{ds-TS}_{\textup{EKF}}$  \\
\hline
\multirow{6}{*}{$\textup{F}_1$ (\%)} 
&1+2& 48.27 $\pm$ 35.61 & 44.19 $\pm$ 34.28 & 83.87 $\pm$ 27.53\\ 
&1+3& 49.47 $\pm$ 35.35 & 48.73 $\pm$ 34.48 & 84.83 $\pm$ 26.47\\ 
&1+4& 55.85 $\pm$ 36.10 & 55.47 $\pm$ 35.26 & 89.48 $\pm$ 21.76\\ 
&2+3& 51.41 $\pm$ 35.84 & 48.71 $\pm$ 36.56 & {\bf 81.61 $\pm$ 29.06}\\ 
&2+4& 55.62 $\pm$ 35.51 & 53.37 $\pm$ 35.92 & 84.07 $\pm$ 27.96\\ 
&3+4& 48.00 $\pm$ 33.62 & 48.92 $\pm$ 33.28 & 78.64 $\pm$ 31.44\\
\hline
\multicolumn{2}{|c|}{$\textup{F}_1(1)$ (\%)} & 63.23$\pm$34.22 & 65.18$\pm$ 31.93 & 91.95$\pm$19.43\\
\hline
\multicolumn{2}{|c|}{$\textup{F}_1(0.5)$ (\%)} & 52.62$\pm$ 34.29 & 50.72$\pm$34.04 & 87.10 $\pm$ 21.95\\\hline\hline
\multirow{6}{*}{MAE (ms)}
&1+2& 15.67 $\pm$ 9.79 & 15.94 $\pm$  9.86 &  9.58 $\pm$ 7.30\\ 
&1+3& 16.26 $\pm$ 10.33& 16.00 $\pm$ 10.17 &  9.59 $\pm$ 7.07\\ 
&1+4& 12.89 $\pm$ 9.45 & 12.60 $\pm$  9.65 &  8.65 $\pm$ 6.09\\ 
&2+3& 15.03 $\pm$ 9.50 & 14.82 $\pm$  8.81 &  9.89 $\pm$ 7.92\\ 
&2+4& 12.61 $\pm$ 9.51 & 12.93 $\pm$  9.41 &  9.15 $\pm$ 7.20\\ 
&3+4& 15.49 $\pm$ 9.57 & 14.82 $\pm$  9.01 & 10.52 $\pm$ 8.24\\
\hline
\multicolumn{2}{|c|}{MAE(1) (ms)} & 11.70$\pm$ 9.46& 10.92$\pm$ 8.73 & 7.93 $\pm$ 5.98 \\
\hline
\multicolumn{2}{|c|}{MAE(0.5) (ms)} &13.88 $\pm$ 9.12& 13.55 $\pm$ 8.65 & 8.82 $\pm$ 6.31\\
\hline

\multicolumn{5}{c}{\vspace{-2pt}}\\
\cline{2-5}
\multicolumn{1}{c|}{} & channels & $\textup{ds-TS}_{\textup{PCA}}$ & $\textup{BSS}_{\textup{ICA}}$ & {$\textup{SVD}_{top1}$}  \\
\hline
\multirow{6}{*}{$\textup{F}_1$ (\%)} 
&1+2& 83.43 $\pm$ 27.78 & 40.92 $\pm$ 32.69 & 79.19 $\pm$ 31.56\\
&1+3& 83.61 $\pm$ 27.75 & 38.55 $\pm$ 31.89 & 75.26 $\pm$ 33.85\\ 
&1+4& 90.20 $\pm$ 18.80 & 40.96 $\pm$ 33.21 & 83.57 $\pm$ 28.20\\ 
&2+3& 79.53 $\pm$ 30.65 & 38.33 $\pm$ 31.41 & 75.38 $\pm$ 34.29\\ 
&2+4& 84.19 $\pm$ 27.95 & 38.35 $\pm$ 31.56 & 78.00 $\pm$ 33.55\\ 
&3+4& 79.53 $\pm$ 30.10 & 36.92 $\pm$ 30.73 & 73.45 $\pm$ 36.54\\
\hline
\multicolumn{2}{|c|}{$\textup{F}_1(1)$ (\%)} & 94.42$\pm$13.66 & 39.01$\pm$ 27.31 & 88.31$\pm$25.28\\
\hline
\multicolumn{2}{|c|}{$\textup{F}_1(0.5)$ (\%)} &85.95 $\pm$ 23.13 & 38.50$\pm$29.44 & 80.77$\pm$30.16\\
\hline\hline
\multirow{6}{*}{MAE (ms)}
&1+2& 7.60 $\pm$ 7.57 & 18.91 $\pm$ 9.06 & 8.55 $\pm$ 8.17\\ 
&1+3& 8.28 $\pm$ 8.39 & 19.19 $\pm$ 8.66 & 9.82 $\pm$ 9.10\\ 
&1+4& 6.11 $\pm$ 5.57 & 18.32 $\pm$ 9.92 & 7.80 $\pm$ 8.02\\ 
&2+3& 8.78 $\pm$ 7.81 & 19.32 $\pm$ 8.54 & 10.06 $\pm$ 8.75\\ 
&2+4& 7.98 $\pm$ 7.69 & 19.10 $\pm$ 8.95 & 8.98 $\pm$ 8.55\\ 
&3+4& 8.33 $\pm$ 7.93 & 19.28 $\pm$ 8.76 & 10.15 $\pm$ 9.25 \\
\hline
\multicolumn{2}{|c|}{MAE(1) (ms)} & 5.28$\pm$ 4.01& 16.33$\pm$ 9.75 & 7.00 $\pm$ 7.45 \\
\hline
\multicolumn{2}{|c|}{MAE(0.5) (ms)} &6.80 $\pm$ 6.15& 18.97$\pm$ 8.45 & 8.09$\pm$7.68\\
\hline

\end{tabular}}
\label{table 1}
\end{table}

\begin{table} 
\scriptsize\centering
\caption{The comparison of $\textup{F}_1$ scores for different 3-channel algorithms applied on \texttt{CinC2013}. The subject a54 is removed from the dataset. All data are presented as mean$\pm$standard deviation.}
{
\begin{tabular}{|c|c|c|c|c|c|c|}
\multicolumn{5}{c}{\vspace{-3pt}}\\
\cline{2-7}
\multicolumn{1}{c|}{} & channels & SAVER \cite{LiFraschWu2017} & {Proposed} &$\textup{ds-TS}_\textup{PCA}$ & {$\textup{SVD}_{top1}$} & $\textup{BSS}_\textup{ICA}$   \\
\hline
\multirow{4}{*}{$\textup{F}_1$ (\%)} 
& 1+2+3 & 80.68 $\pm$ 29.52 &{\bf 89.70 $\pm$ 23.46} &86.33 $\pm$ 25.81 & 77.05 $\pm$ 34.90 & 22.31 $\pm$ 16.50 \\
& 1+2+4 & 84.53 $\pm$ 26.22 &{\bf 93.91 $\pm$ 14.83} &87.24 $\pm$ 25.46 & 75.24 $\pm$ 35.33 & 25.74 $\pm$ 23.69 \\
& 1+3+4 & 83.52 $\pm$ 27.98 &{\bf 93.62 $\pm$ 14.57} &88.08 $\pm$ 23.47 & 80.73 $\pm$ 31.86 & 26.66 $\pm$ 24.02 \\
& 2+3+4 & 77.51 $\pm$ 31.47 &{\bf 84.82 $\pm$ 27.31} &80.28 $\pm$ 31.23 & 78.01 $\pm$ 34.01 & 24.51 $\pm$ 21.04 \\

\hline
\multicolumn{2}{|c|}{$\textup{F}_1(1)$ (\%)}&91.20 $\pm$ 18.47 & {\bf 95.32$\pm$ 13.75} & 89.64 $\pm$ 22.94 & 87.16$\pm$ 26.77 & 33.83$\pm$30.07 \\
\hline
\multicolumn{2}{|c|}{$\textup{F}_1(0.5)$ (\%)} & 82.79$\pm$26.67 & {\bf 91.90$\pm$ 17.38} & 86.81$\pm$25.06 &79.50 $\pm$ 32.15& 24.09 $\pm$ 19.23\\
\hline\hline
\multirow{4}{*}{MAE (ms)}
& 1+2+3 & 7.15 $\pm$ 6.29 &{\bf 6.75 $\pm$ 7.28} &6.11 $\pm$ 5.05 &9.74 $\pm$ 8.61 & 21.95 $\pm$ 5.28 \\
& 1+2+4 & 6.44 $\pm$ 6.13 &{\bf 5.64 $\pm$ 5.11} &6.48 $\pm$ 6.76 &9.68 $\pm$ 9.56 & 21.38 $\pm$ 7.02 \\
& 1+3+4 & 6.73 $\pm$ 6.43 &{\bf 5.47 $\pm$ 5.09} &6.11 $\pm$ 5.79 &8.87 $\pm$ 8.48 & 20.25 $\pm$ 7.47 \\
& 2+3+4 & 7.97 $\pm$ 7.08 &{\bf 7.82 $\pm$ 7.23} &8.12 $\pm$ 8.00 &9.39 $\pm$ 9.01 & 21.61 $\pm$ 6.37 \\
\hline
\multicolumn{2}{|c|}{MAE(1) (ms)}&  {\bf 5.22 $\pm$ 3.87} &   5.41$\pm$ 5.22 & 5.77 $\pm$ 5.23 & 7.58 $\pm$ 7.85 & 22.27 $\pm$ 8.39 \\
\hline
\multicolumn{2}{|c|}{MAE(0.5) (ms)} & 6.67 $\pm$ 5.68 &  {\bf 5.80$\pm$ 5.36} & 6.32$\pm$6.02 & 8.79$\pm$8.24 & 21.76 $\pm$ 5.85\\
\hline
\end{tabular}}
\label{table 3}
\end{table}

\begin{table} 
\scriptsize\centering
\caption{The comparison of $\textup{F}_1$ and MAE for different algorithms applied to \texttt{CinC2013} when we only consider single channel. The subject a54 is removed from the dataset. All data are presented as mean$\pm$standard deviation.}
{
\begin{tabular}{|c|c|c|c|c|}
\multicolumn{5}{c}{\vspace{-3pt}}\\
\cline{2-5}
\multicolumn{1}{c|}{} & channel & \cite{SuWu2017} & {Proposed} &    {$\textup{SVD}_{top1}$}\\
\hline
\multirow{4}{*}{$\textup{F}_1$ (\%)}
&1& 68.38 $\pm$ 33.12 & 66.29 $\pm$ 36.76 & 63.57 $\pm$ 35.50\\ 
&2& 74.35 $\pm$ 29.96 & 75.63 $\pm$ 30.96 & 72.47 $\pm$ 33.17\\ 
&3& 64.10 $\pm$ 33.58 & 65.75 $\pm$ 35.09 & 62.36 $\pm$ 36.33\\ 
&4& 75.68 $\pm$ 29.64 & 74.70 $\pm$ 32.40 & 71.92 $\pm$ 34.12\\ 

\hline
\multicolumn{2}{|c|}{$\textup{F}_1(1)$ (\%)} & 87.01$\pm$22.06 & 88.05$\pm$ 23.08 & 83.72 $\pm$26.27\\
\hline
\multicolumn{2}{|c|}{$\textup{F}_1(0.5)$ (\%)} & 74.06 $\pm$ 27.63 & 72.67$\pm$ 29.92 & 69.62$\pm$31.83\\
\hline\hline
\multirow{4}{*}{MAE (ms)}
&1& 11.33 $\pm$ 9.31 & 11.79 $\pm$ 9.62 & 12.05$\pm$9.69\\
&2& 9.86 $\pm$ 7.99 & 8.97 $\pm$  7.77 & 10.53$\pm$8.84\\ 
&3& 12.17 $\pm$ 9.04 & 12.49 $\pm$ 9.48 & 12.95$\pm$9.43\\ 
&4&  8.98 $\pm$ 7.95 &  9.06 $\pm$  8.50 & 9.94$\pm$8.81\\ 

\hline
\multicolumn{2}{|c|}{MAE(1) (ms)} & 5.88$\pm$ 5.14&  6.08$\pm$5.41 & 7.09$\pm$7.21 \\
\hline
\multicolumn{2}{|c|}{MAE(0.5) (ms)} & 9.56 $\pm$ 6.86 & 9.69$\pm$ 7.37 &10.57 $\pm$ 7.87\\
\hline
\multicolumn{5}{c}{\vspace{-2pt}}\\
\cline{2-5}
\multicolumn{1}{c|}{} &channel & $\textup{ds-TS}_{\textup{C}}$ & $\textup{ds-TS}_{\textup{PCA}}$ & $\textup{ds-TS}_{\textup{EKF}}$  \\
\hline
\multirow{4}{*}{$\textup{F}_1$ (\%)} 
&1 & 61.17 $\pm$ 37.63 & 62.64 $\pm$ 38.08 & 61.33 $\pm$ 35.34\\ 
&2 & 64.47 $\pm$ 36.95 & 64.57 $\pm$ 36.31 & 67.67 $\pm$ 35.88\\ 
&3 & 55.55 $\pm$ 37.02 & 56.97 $\pm$ 36.78 & 57.32 $\pm$ 37.54\\ 
&4 & 67.45 $\pm$ 36.34 & 69.08 $\pm$ 35.23 & 68.24 $\pm$ 35.06\\ 
\hline
\multicolumn{2}{|c|}{$\textup{F}_1(1)$ (\%)} & 82.73$\pm$29.32 & 84.47$\pm$ 27.75 & 82.31$\pm$28.05\\
\hline
\multicolumn{2}{|c|}{$\textup{F}_1(0.5)$ (\%)} & 68.53 $\pm$ 33.13 & 68.36$\pm$ 33.16 & 70.81$\pm$32.13\\
\hline\hline
\multirow{4}{*}{MAE (ms)}
&1& 12.88 $\pm$ 10.23 & 12.29 $\pm$ 9.68 & 14.63 $\pm$ 8.27\\ 
&2& 11.16 $\pm$  8.64 & 10.72 $\pm$ 8.30 & 12.22 $\pm$ 8.21\\ 
&3& 13.69 $\pm$  9.55 & 13.25 $\pm$ 9.67 & 14.64 $\pm$ 8.43\\ 
&4&  9.30 $\pm$  8.26 &  8.94 $\pm$ 8.00 & 12.15 $\pm$ 8.03\\ 

\hline
\multicolumn{2}{|c|}{MAE(1) (ms)} & 6.84$\pm$ 6.91 &  6.64$\pm$6.61 & 9.96 $\pm$ 7.24 \\
\hline
\multicolumn{2}{|c|}{MAE(0.5) (ms)} & 11.37 $\pm$ 8.58 & 11.35$\pm$ 8.52 & 12.53$\pm$7.89\\
\hline
\multicolumn{5}{c}{\vspace{-2pt}}\\
\cline{2-5}
\multicolumn{1}{c|}{} & channel & $\textup{ds-AM}_{\textup{LMS}}$ & $\textup{ds-AM}_{\textup{RLS}}$ & $\textup{ds-AM}_{\textup{ESN}}$  \\
\hline
\multirow{4}{*}{$\textup{F}_1$ (\%)} 
&1 & 43.04 $\pm$ 31.50 & 45.76 $\pm$ 34.42 & 47.24 $\pm$ 33.91\\ 
&2 & 50.87 $\pm$ 32.80 & 53.57 $\pm$ 35.21 & 51.73 $\pm$ 35.09\\ 
&2 & 43.67 $\pm$ 33.51 & 45.30 $\pm$ 36.43 & 45.15 $\pm$ 34.98\\ 
&4 & 49.35 $\pm$ 34.23 & 51.78 $\pm$ 35.27 & 50.19 $\pm$ 34.04\\ 

\hline
\multicolumn{2}{|c|}{$\textup{F}_1(1)$ (\%)} & 63.79$\pm$31.80 & 67.34$\pm$ 33.10 & 68.32$\pm$31.27\\
\hline
\multicolumn{2}{|c|}{$\textup{F}_1(0.5)$ (\%)} & 47.50 $\pm$ 30.16 & 54.43$\pm$ 33.62 & 53.79$\pm$33.06\\
\hline\hline
\multirow{4}{*}{MAE (ms)}
&1 & 15.46 $\pm$ 9.18 & 14.37 $\pm$ 9.65 & 14.23 $\pm$ 9.35\\ 
&2 & 12.94 $\pm$ 8.74 & 11.45 $\pm$ 8.17 & 12.67 $\pm$ 8.72\\ 
&3 & 15.23 $\pm$ 8.11 & 14.95 $\pm$ 9.34 & 13.63 $\pm$ 8.38\\ 
&4 & 13.33 $\pm$ 9.00 & 12.14 $\pm$ 9.38 & 12.52 $\pm$ 9.64\\ 

\hline
\multicolumn{2}{|c|}{MAE(1) (ms)} & 10.21$\pm$ 8.70 &  8.92$\pm$7.58 & 8.13 $\pm$ 6.67 \\
\hline
\multicolumn{2}{|c|}{MAE(0.5) (ms)} & 13.76 $\pm$ 7.26& 12.25$\pm$ 8.25 & 12.38$\pm$8.18\\
\hline
\end{tabular}
}
\label{table 2}
\end{table}

\begin{table} 
\scriptsize\centering
\caption{{ The comparison of $\textup{F}_1$ and MAE of Proposed algorithms applied to \texttt{CinC2013} with three matching windows, 10 ms, 25 ms and 50 ms, and only two channels are considered. The subject a54 is removed from the dataset. All data are presented as mean$\pm$standard deviation.} \label{Table:Different Windows}}
{
\begin{tabular}{|c|c|c|c|c|}
\multicolumn{5}{c}{\vspace{-3pt}}\\
\cline{2-5}
\multicolumn{1}{c|}{} & channels & 10 ms & 25 ms & 50 ms \\
\hline
\multirow{6}{*}{$\textup{F}_1$ (\%)} & 1+2 &77.45$\pm$ 29.80 & 82.77 $\pm$ 27.61 &  86.62 $\pm$ 23.54 \\
& 1+3 & 81.15 $\pm$ 26.55 & 86.27$\pm$24.51 & 89.72 $\pm$ 21.76 \\
& 1+4 & 84.15 $\pm$ 20.65 & 89.95$\pm$16.98 & 93.21 $\pm$ 14.31\\
& 2+3 & 69.13 $\pm$ 34.83 & 75.04$\pm$34.20 & 79.25 $\pm$ 31.75 \\
& 2+4 & 78.30 $\pm$28.69 & 83.21$\pm$27.13 & 87.32$\pm$ 24.44 \\
& 3+4 & 75.08 $\pm$ 32.21 & 78.67$\pm$31.09& 82.60 $\pm$ 28.03 \\
\hline
\multicolumn{2}{|c|}{$\textup{F}_1(1)$ (\%)} &90.67$\pm$14.83 & 94.05$\pm$ 13.16& 96.31$\pm$ 10.93 \\
\hline
\multicolumn{2}{|c|}{$\textup{F}_1(0.5)$ (\%)} &79.78$\pm$ 25.05 & 84.44 $\pm$ 23.94 &  87.94 $\pm$ 21.30 \\
\hline
\hline
\multirow{6}{*}{MAE (ms)} & 1+2 & 3.39$\pm$ 2.13 & 4.50$\pm$3.26 & 6.86 $\pm$ 6.55 \\
& 1+3 & 3.58 $\pm$ 2.09 & 4.43$\pm$3.01 & 6.43 $\pm$ 6.37 \\
& 1+4 & 3.36 $\pm$ 2.15 & 4.04$\pm$2.60 & 5.44 $\pm$ 4.18 \\
& 2+3 & 3.82 $\pm$ 1.99 & 5.54$\pm$3.69 & 8.61 $\pm$ 7.51\\
& 2+4 & 3.45 $\pm$ 1.90 & 4.70$\pm$3.18 & 7.11$\pm$ 6.77 \\
& 3+4 & 3.27 $\pm$ 2.05 & 4.50$\pm$3.39 & 7.27 $\pm$ 7.29 \\
\hline
\multicolumn{2}{|c|}{MAE(1) (ms)} & 3.21$\pm$1.99 & 3.92$\pm$2.50 &  4.93$\pm$3.64 \\
\hline
\multicolumn{2}{|c|}{MAE(0.5) (ms)}& 3.38 $\pm$ 1.85 & 4.36$\pm$2.54 & 6.37 $\pm$ 5.71 \\
\hline

\end{tabular}}

\label{table 4}
\end{table}  

\end{document}